%% file: main.tex
\documentclass[10pt,twocolumn]{article}

\usepackage[utf8]{inputenc}
\usepackage[T1]{fontenc}
\usepackage[letterpaper,margin=0.75in,columnsep=0.25in]{geometry}
\usepackage{amsmath,amssymb,amsfonts}
\usepackage{graphicx}
\usepackage{booktabs}      % \toprule/\midrule/\bottomrule -- matches pandas to_latex() output
\usepackage{caption}
\usepackage{subcaption}
\usepackage[colorlinks=true,linkcolor=blue,citecolor=blue,urlcolor=blue]{hyperref}
\usepackage[numbers,sort&compress]{natbib}
\usepackage{xcolor}
\usepackage{multirow}
\usepackage{algorithm}
\usepackage{algpseudocode}
\usepackage{tikz}
\usetikzlibrary{arrows.meta,positioning,shapes.geometric,fit,backgrounds}
\usepackage{changepage} % \adjustwidth, for the wider-margin arXiv-style abstract block

\usepackage{fancyhdr}
\newcommand{\pce}{PCE}
\newcommand{\dpce}{dPCE}

\graphicspath{{./figures/}}

\title{Large-Scale Streaming Portfolio Optimization with Pauli Correlation Encoding}
\author{Vicente P. Soloviev, Josh Kirsopp}

\begin{document}

\twocolumn[
  \begin{center}
    {\LARGE\textbf{Large-Scale Streaming Portfolio Optimization with Pauli
    Correlation Encoding}}
    \par\vspace{2ex}
    \begin{tabular}{c@{\hspace{3em}}c}
      Vicente P. Soloviev\textsuperscript{*} & Josh Kirsopp \\
      Fujitsu Research of Europe Ltd. & Fujitsu Research of Europe Ltd. \\
      Pozuelo de Alarc\'on, 28224 Madrid, Spain & Slough SL1 2BE, UK \\
      \texttt{vicente.perezsoloviev@fujitsu.com} & \\
    \end{tabular}
    \par\vspace{1ex}
    {\small \textsuperscript{*}Corresponding author.}
  \end{center}
  \vspace{2em}
  \begin{adjustwidth}{0.5in}{0.5in}
    \begin{center}
      {\large\textbf{Abstract}}
    \end{center}
    \small
    \input{sections/00_abstract}
  \end{adjustwidth}
  \vspace{2em}
]

\section{Introduction}
\label{sec:introduction}
\input{sections/01_introduction}

\section{Methodology}
\label{sec:methodology}
\input{sections/02_methodology}

\section{Portfolio Optimization via \pce{} and \dpce{}}
\label{sec:finance-mapping}
\input{sections/03_finance_mapping}

\section{Results}
\label{sec:results}
\input{sections/04_results}

\section{Conclusion and Discussion}
\label{sec:conclusion}
\input{sections/05_conclusion}

\bibliographystyle{unsrtnat}
\bibliography{references}

\appendix
\input{sections/06_appendix_datasets}

\end{document}

%% file: sections/00_abstract.tex
Classical portfolio construction over large asset universes is a combinatorial selection problem
that existing heuristics do not solve well at scale. Pauli Correlation Encoding (\pce{})
addresses this by representing the underlying clustering problem with far fewer qubits than one
per asset, and a prior study showed it is competitive with classical methods on a single
historical snapshot. This paper moves beyond that static test to a streaming deployment, in
which the correlation structure and the portfolio are both re-estimated on a rolling basis, as a
live rebalancing system would actually operate. We introduce two streaming variants -- \pce{}
itself and a joint multi-cut variant, \dpce{} -- and evaluate both against a Buy~\&~Hold baseline
across window lengths, portfolio sizes up to several thousand assets, risk-aversion and
transaction-cost settings, and measurement-shot budgets. Streaming \pce{} and \dpce{}
consistently outperform the passive baseline, including during a market downturn where both lose
money but \pce{} loses markedly less -- evidence that the method's value is not confined to
rising markets. \pce{} also proves the more practical choice at scale, since \dpce{}'s resource
requirements and its sensitivity to rebalancing frequency both grow faster with portfolio size.
We further report which operational choices -- rebalancing cadence, shot budget, and
transaction-cost awareness -- matter most for a real deployment, and which matter less than
commonly assumed.

%% file: sections/01_introduction.tex
Selecting a portfolio from a large universe of candidate assets is, at its core, a combinatorial
problem: choosing which subset to hold, out of exponentially many possibilities, to balance
diversification against concentration in the most promising opportunities.
\citet{markowitz1952portfolio} formalized the trade-off between expected return and risk that any
such selection must respect, but exact mean-variance optimization over a discrete, large
candidate set is intractable: cardinality-constrained portfolio selection is NP-hard in
general~\citep{egger2020quantum}, and the heuristics the finance industry actually deploys to
sidestep that intractability trade solution
quality for tractability rather than preserving both. Hierarchical clustering on an
asset-correlation matrix~\citep{lopezdeprado2016hrp}, risk-parity allocation, and spectral
graph partitioning are all variations on the same idea -- replace the discrete selection problem
with a clustering step that is cheap to compute -- and all of them degrade as the candidate
universe grows into the thousands of names, precisely because the approximations that keep them
cheap are what erode the diversification structure the method was trying to recover in the first
place. This is the specific gap that motivates looking at quantum-inspired combinatorial
encodings for this problem: not that classical methods cannot run at scale, but that the ones
cheap enough to run at scale do not preserve the quality of the answer they return.

Quantum approaches to combinatorial optimization have historically faced the opposite problem.
The Quantum Approximate Optimization Algorithm~\citep{farhi2014qaoa} and quantum-annealing
formulations of portfolio selection~\citep{rosenberg2016quantum,orus2019quantum} typically
dedicate one qubit per decision variable, which makes them competitive on small instances but
puts problem sizes relevant to real portfolio universes -- hundreds to thousands of assets --
far outside the reach of near-term hardware, a gap documented across several quantum-finance
surveys~\citep{egger2020quantum,herman2023quantum}. Pauli Correlation Encoding (\pce{}),
introduced by \citet{sciorilli2024pce}, closes this gap from the encoding side rather than the
hardware side: an $n$-variable graph-cutting instance is represented using only $O(\log n)$
qubits, by encoding each decision variable as a function of a high-order Pauli correlator
instead of giving it a dedicated qubit. A parameterized circuit is optimized, classically, so
that its measured Pauli-correlator expectation values encode a high-quality graph cut; because
the qubit count grows only logarithmically with problem size, \citeauthor{sciorilli2024pce}
validated the approach on graph-cutting instances up to several hundred variables, including
runs on trapped-ion hardware -- scales that remain out of reach for one-qubit-per-variable
encodings on the same devices.

That validation, however, concerns generic graph-cutting instances, not portfolios specifically,
and it is static: a single problem instance, solved once. \citet{soloviev2026pce} closed that
second gap, applying \pce{} directly to portfolio construction -- assets become nodes in a
correlation graph, \pce{}-based graph cutting groups them into diversified clusters, and one
representative asset per cluster becomes the portfolio -- and showing the resulting portfolios
are competitive with classical baselines on real market data. That study is, by design, a static one: one historical correlation snapshot, one graph cut, one portfolio, evaluated
once on held-out data.

Since its introduction, \pce{} has been extended and benchmarked rapidly across a growing range
of settings beyond this original graph-cutting demonstration. \citet{padin2026progressive}
extend it to constrained combinatorial optimization via a progressive-binarization continuation
scheme; \citet{soloviev2026cpce} generalize the encoding itself from discrete to continuous-domain
data for classical-correlation learning; \citet{alonso2026benchmark} and \citet{reeves2026mis}
benchmark \pce{}'s performance and ansatz choices across several combinatorial problem classes
beyond graph cutting, including the Maximum Independent Set problem, while
\citet{sharma2025comparative} place it alongside other qubit-compression and quantum-optimization
techniques in a broader comparative study of NP-hard benchmark problems;
\citet{yoshioka2026powerdemand} apply it to a second large-scale real-world portfolio problem --
power-demand allocation -- independently of the finance application studied here; and
\citet{docarmo2025warmpce} show that warm-starting \pce{} with a classically-derived bias
improves solution quality on the Traveling Salesman Problem, a precedent for the warm-starting
mechanism we evaluate in a streaming context in \S\ref{sec:results-warmstart}. This emerging body
of work establishes \pce{} as an actively-studied encoding rather than a single-paper result, but,
to our knowledge, none of it addresses the streaming, repeatedly-rebalanced deployment setting
that is this paper's focus.

\citeauthor{soloviev2026pce}'s static study, however, remains the right first test of \pce{} for
portfolios specifically, and it leaves open exactly
the question a deployed system cannot avoid answering: a live portfolio manager does not compute
one portfolio and hold it forever. The correlation structure between assets drifts continuously,
and any real deployment must periodically re-estimate it and re-cluster on a rolling basis --
what we refer to throughout this paper as streaming or real-time operation,
meaning not that any individual computation must complete within milliseconds, but that the
whole pipeline -- data ingestion, graph construction, quantum (or quantum-simulated)
optimization, and portfolio selection -- must run repeatedly, on a fixed rebalancing cadence,
indefinitely.

Moving from a static demonstration to this streaming setting changes what needs to be validated,
and that shift is the subject of this paper. Every design choice that a static study fixes once
-- the lookback window used to estimate correlations, the number of measurement shots spent per
optimization, whether transaction costs are priced into asset selection, how aggressively the
portfolio is rebalanced -- recurs at every rebalance in a streaming deployment, and its cost and
benefit compound over the life of the system rather than being paid a single time. Streaming
deployment also raises a question a static study cannot pose at all: whether solution quality
and computational cost remain favorable as the live universe grows, when every rebalance,
not just one, issues a fresh optimization call. The remainder of this paper studies two
streaming variants of \pce{} -- an iterative recursive-bipartition scheme and a joint multi-cut
scheme, \dpce{} -- under exactly these conditions, on real market data, across the parameters a
live deployment actually has to choose rather than fix in advance.

The remainder of the paper is organized as follows. Section~\ref{sec:methodology} formalizes the
graph-cutting problem underlying both algorithms, derives the $O(\log n)$ qubit requirement, and
specifies \pce{} and \dpce{} as two concrete streaming schemes. Section~\ref{sec:finance-mapping}
maps that generic formulation onto the portfolio problem specifically: how the market graph is
built from correlation data, how clusters become a tradable portfolio, and how transaction costs,
a tunable risk-aversion profile, the rolling-window mechanics, and the classical optimizer and
ansatz choice all enter the pipeline, each grounded in the corresponding piece of classical
portfolio theory. Section~\ref{sec:results} reports the empirical results -- streaming \pce{} and
\dpce{} performance across window length, risk-aversion, and transaction-cost settings; a direct
comparison between the two algorithms; the effect of rebalance cadence, warm-starting, and shot
budget; and scalability up to $N=5{,}000$ candidate assets. Section~\ref{sec:conclusion}
summarizes the findings and discusses their implications for deploying \pce{}-based optimization
in practice. Appendix~\ref{app:data} documents the datasets used throughout.

%% file: sections/02_methodology.tex
This section introduces \pce{} and \dpce{} as algorithms on an arbitrary weighted graph, and the
loss function each optimizes. We deliberately describe both in graph-theoretic terms --
graph cutting rather than the more specific ``MaxCut'' label -- since the portfolio
application in \S\ref{sec:finance-mapping} uses the general weighted-cut formulation throughout,
and the loss-function modifications needed to encode transaction costs and a tunable risk
profile (also \S\ref{sec:finance-mapping}) are finance-specific choices layered on top of this
general algorithm, not part of it.

\subsection{The graph-cutting problem}
Given a weighted graph $G=(V,E,w)$ with $|V|=n$, a cut is a bipartition (or, for $k>2$
clusters, a $k$-partition) of $V$ that assigns each node a discrete label. Its value is the sum
of edge weights crossing between differently-labeled nodes,
\begin{equation}
  C(b) = \sum_{(u,v)\in E} w_{uv}\,\mathbb{1}[b_u \neq b_v],
  \label{eq:cut-value}
\end{equation}
and the graph-cutting problem is to find the labeling $b$ maximizing $C(b)$. This is NP-hard in
general; \pce{} and \dpce{} both approximate it by relaxing the discrete labels to continuous
variables optimized by a parameterized quantum circuit, then rounding back to a discrete
labeling at the end.

\subsection{Qubit requirement}
\label{sec:qubit-requirement}
\pce{} represents an $n$-node problem using $m$ qubits, where $m$ is the smallest integer such
that
\begin{equation}
  3 \sum_{k=1}^{\text{order}} \binom{m}{k} \;\geq\; n ,
  \label{eq:calculate_qubits}
\end{equation}
i.e.\ enough Pauli correlators of order $k\leq\texttt{order}$ ($X$, $Y$, $Z$ on each of the
$\binom{m}{k}$ possible $k$-qubit subsets, for $k=1,\dots,\text{order}$) must exist to give every
one of the $n$ decision variables its own correlator, rather than its own qubit. \texttt{order}
trades qubit count against correlator degree: raising it shrinks $m$ combinatorially but
increases the number and weight of higher-order correlator terms the ansatz must express.
We fix \texttt{order} per universe size $N$ to the smallest value keeping
qubit count within a simulator-feasible budget, following the progression used throughout this
paper ($\text{order}=2$ at $N\in\{50,100\}$, $\text{order}=3$ at $N\in\{200,300,1{,}601\}$,
$\text{order}=4$ at $N=5{,}000$) rather than optimizing qubit-efficiency at every scale, since
the qubit count that results is already comfortable at every tested $N$
(\S\ref{sec:results-scalability}): $6$ qubits at $N=50$, $8$ at $N\in\{100,200\}$, $9$ at
$N=300$, and $15$ at both $N=1{,}601$ and $N=5{,}000$.

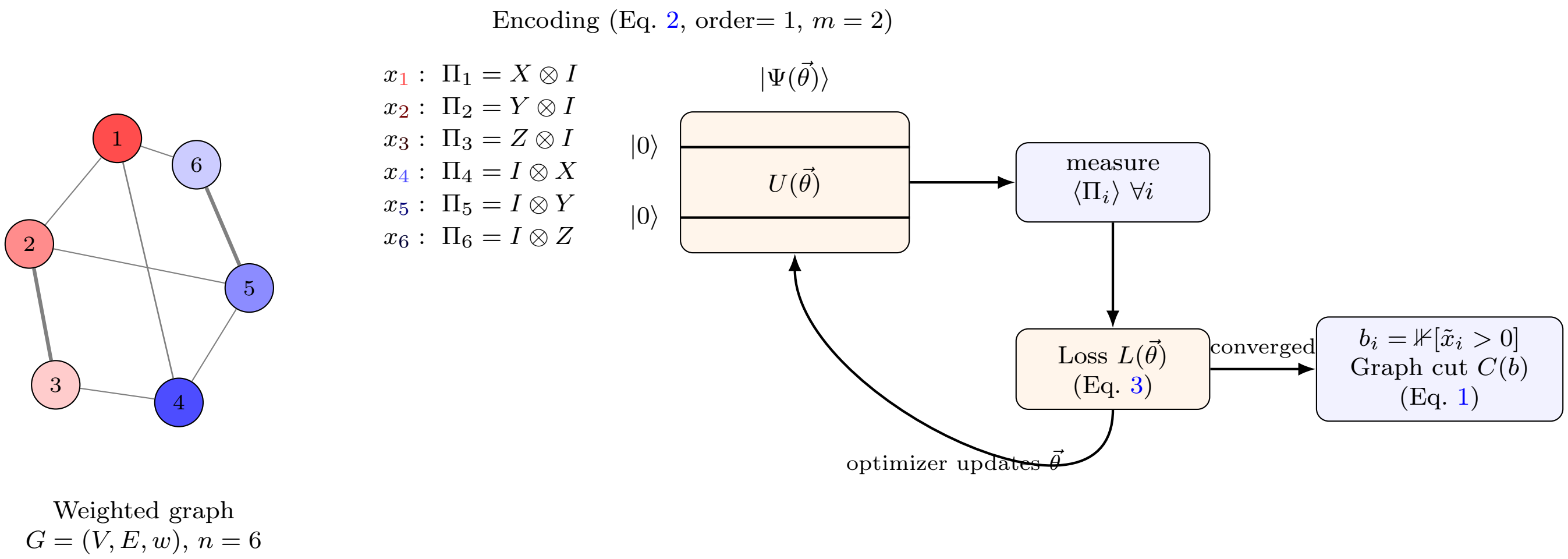
\begin{figure*}[t]
  \centering
  \begin{tikzpicture}[
      every node/.style={font=\footnotesize},
      gnode/.style={circle, draw, minimum size=5.5mm, inner sep=0pt, font=\scriptsize},
      box/.style={draw, rounded corners, align=center, minimum height=9mm, minimum width=22mm,
                  font=\footnotesize, fill=blue!5},
      loopbox/.style={box, fill=orange!8},
      arr/.style={-{Latex[length=2mm]}, thick},
    ]
    % --- Worked example graph: n=6 nodes, order=1, m=2 qubits ---
    \node[gnode, fill=red!70]   (n1) at (0, 1.5)   {1};
    \node[gnode, fill=red!45]   (n2) at (-1.0, 0.3) {2};
    \node[gnode, fill=red!20]   (n3) at (-0.7,-1.3) {3};
    \node[gnode, fill=blue!70]  (n4) at (0.7,-1.5)  {4};
    \node[gnode, fill=blue!45]  (n5) at (1.5,-0.2)  {5};
    \node[gnode, fill=blue!20]  (n6) at (0.9, 1.2)  {6};
    \draw[gray] (n1)--(n2);
    \draw[gray,line width=1.2pt] (n2)--(n3);
    \draw[gray] (n3)--(n4);
    \draw[gray] (n4)--(n5);
    \draw[gray,line width=1.2pt] (n5)--(n6);
    \draw[gray] (n6)--(n1);
    \draw[gray] (n2)--(n5);
    \draw[gray,line width=0.5pt] (n1)--(n4);
    \node[below=2mm, align=center] at (0.3,-2.3) {Weighted graph\\$G=(V,E,w)$, $n=6$};

    % --- Encoding list, color-matched to graph nodes ---
    \node[align=left, anchor=west] (enc) at (2.9, 1.3) {
      $x_{\textcolor{red!70}{1}}:\ \Pi_1 = X\otimes I$\\[1pt]
      $x_{\textcolor{red!45!black}{2}}:\ \Pi_2 = Y\otimes I$\\[1pt]
      $x_{\textcolor{red!20!black}{3}}:\ \Pi_3 = Z\otimes I$\\[1pt]
      $x_{\textcolor{blue!70}{4}}:\ \Pi_4 = I\otimes X$\\[1pt]
      $x_{\textcolor{blue!45!black}{5}}:\ \Pi_5 = I\otimes Y$\\[1pt]
      $x_{\textcolor{blue!20!black}{6}}:\ \Pi_6 = I\otimes Z$
    };
    \node[above=1mm of enc, anchor=south west] {Encoding (Eq.~\ref{eq:calculate_qubits},
      order$=1$, $m=2$)};

    % --- Circuit box (top row) ---
    \node[loopbox, minimum width=26mm, minimum height=16mm] (circuit) at (7.7, 1.0) {};
    \draw[thick] ([yshift=4mm]circuit.west) -- ([yshift=4mm]circuit.east);
    \draw[thick] ([yshift=-4mm]circuit.west) -- ([yshift=-4mm]circuit.east);
    \node[left=1mm of circuit, anchor=east] at ([yshift=4mm]circuit.west) {$|0\rangle$};
    \node[left=1mm of circuit, anchor=east] at ([yshift=-4mm]circuit.west) {$|0\rangle$};
    \node at (circuit) {$U(\vec\theta)$};
    \node[above=1mm of circuit] {$|\Psi(\vec\theta)\rangle$};

    \node[box, right=12mm of circuit, minimum width=22mm] (measure)
      {measure\\$\langle\Pi_i\rangle\ \forall i$};
    \draw[arr] (circuit) -- (measure);

    % --- Bottom row: loss below measure, output right of loss ---
    \node[loopbox, below=12mm of measure, minimum width=22mm] (loss)
      {Loss $L(\vec\theta)$\\(Eq.~\ref{eq:pce-loss})};
    \draw[arr] (measure) -- (loss);

    \node[box, right=12mm of loss, minimum width=28mm] (output)
      {$b_i=\mathbb{1}[\tilde x_i>0]$\\Graph cut $C(b)$\\(Eq.~\ref{eq:cut-value})};
    \draw[arr] (loss) -- node[above,font=\scriptsize]{converged} (output);

    \draw[arr] (loss.south) .. controls +(0,-16mm) and +(0,-16mm) ..
      node[below, font=\scriptsize, pos=0.5, align=center] {optimizer updates $\vec\theta$}
      (circuit.south);
  \end{tikzpicture}
  \caption{\pce{}/\dpce{} pipeline, worked through a concrete example: a 6-node weighted graph is
  encoded onto $m=2$ qubits by assigning each node its own Pauli correlator $\Pi_i$ (node color
  matches its correlator's label; this order$=1$, $m=2$ instance of
  Eq.~\eqref{eq:calculate_qubits} is a simplified illustration -- \S\ref{sec:qubit-requirement}'s
  experiments use order $\in\{2,3\}$). A parameterized circuit $U(\vec\theta)$ is measured to
  obtain each $\langle\Pi_i\rangle$; a classical optimizer (orange, looped) updates $\vec\theta$ to
  minimize the cut-based loss until convergence, at which point the converged correlator values
  are thresholded into a discrete labeling $b$ and the realized graph cut $C(b)$ is evaluated.
  \pce{} and \dpce{} share this pipeline and differ only in how many labels are encoded per call
  and in the loss function used (\S\ref{sec:pce-algorithm}--\ref{sec:dpce-algorithm}).}
  \label{fig:pipeline}
\end{figure*}

Figure~\ref{fig:pipeline} summarizes this pipeline end to end: graph construction, qubit
encoding, the classical-quantum optimization loop, and final discretization into a graph cut.
\pce{} and \dpce{} instantiate the same pipeline; they differ in how many cluster labels each
quantum call encodes, and consequently in the loss function (\S\ref{sec:base-loss}--\ref{sec:dpce-algorithm}).

\subsection{\pce{} loss function}
\label{sec:base-loss}
Given the regularized (tanh-relaxed) correlator values $\tilde x_i = \tanh(\alpha\,
\text{result}_i)$, $\alpha = m^{\,\text{order}/2}$ by default, the optimizer minimizes
\begin{equation}
  L_{\text{\pce{}}} \;=\; \underbrace{\sum_{(u,v)\in E} w_{uv}\, \tilde x_u \tilde x_v}_{\text{cut term}}
  \;+\; \underbrace{\beta \, v \left(\frac{1}{n}\sum_i \tilde x_i^2\right)^{\!2}}_{\text{balance regularizer}},
  \label{eq:pce-loss}
\end{equation}
with $v = |E|/2 + (n-1)/4$ and $\beta=0.5$. Minimizing the cut term pushes
$\tilde x_u \tilde x_v \to -1$ (opposite sides) for high-weight edges -- the standard continuous
relaxation of Eq.~\eqref{eq:cut-value} for $k=2$ -- while the regularizer penalizes unbalanced
partitions (all $\tilde x_i$ pinned to the same sign). The realized (discrete) objective reported
throughout this paper is the cut value $C(b)$ of Eq.~\eqref{eq:cut-value}, evaluated on the
thresholded labels $b_i = \mathbb{1}[\tilde x_i>0]$, not the continuous relaxation
\eqref{eq:pce-loss} itself.

\subsection{\pce{}: iterative recursive bipartition}
\label{sec:pce-algorithm}
To produce $k$ clusters from an $n$-node graph, \pce{} applies Eq.~\eqref{eq:pce-loss} as a
binary ($b_i\in\{0,1\}$) split, then recursively re-splits whichever resulting cluster is
currently largest, until $k$ clusters exist -- Algorithm~\ref{alg:pce}. Each of the $k-1$ splits
is an independent quantum optimization call on the (shrinking) sub-cluster's own induced
subgraph, re-applying Eq.~\eqref{eq:calculate_qubits}--\eqref{eq:pce-loss} to that subgraph's own
node count. Because later splits operate on smaller subgraphs, their qubit requirement is the
same or lower than the first split's, so the first (full-graph) split's qubit count upper-bounds
every call in the tree.

\begin{algorithm}[t]
\caption{\pce{}: iterative recursive bipartition}
\label{alg:pce}
\begin{algorithmic}[1]
\Require graph $G=(V,E,w)$, target cluster count $k$
\State $\mathcal{P} \gets \{V\}$ \Comment{start with one cluster containing all nodes}
\While{$|\mathcal{P}| < k$}
  \State $S \gets$ largest cluster in $\mathcal{P}$
  \If{$|S| \leq 2$}
    \State \textbf{break} \Comment{nothing useful left to split}
  \EndIf
  \State $G_S \gets$ induced subgraph of $G$ on $S$
  \State $m \gets$ qubits via Eq.~\eqref{eq:calculate_qubits} applied to $|S|$
  \State $b \gets$ optimize Eq.~\eqref{eq:pce-loss} on $G_S$ with $m$ qubits
  \State $S_0, S_1 \gets \{i\in S : b_i=0\}, \{i\in S : b_i=1\}$
  \State $\mathcal{P} \gets (\mathcal{P}\setminus\{S\}) \cup \{S_0, S_1\}$
\EndWhile
\State \Return $\mathcal{P}$
\end{algorithmic}
\end{algorithm}

\subsection{\dpce{}: joint multi-cut}
\label{sec:dpce-algorithm}
\dpce{} instead encodes all $k$ clusters directly in a single call. Each node $i$ is given
$\kappa = \lceil \log_2 k \rceil$ binary variables $b_i^1,\dots,b_i^\kappa$ (its cluster label in
binary), by replicating the original $n$-node graph into $\kappa$ structurally-identical
``layers'' -- a single joint graph of $\kappa n$ nodes, to which Eq.~\eqref{eq:calculate_qubits}
is applied directly. Writing $\tilde x_i^l$ for node $i$'s regularized value on layer $l$, the
optimizer minimizes
\begin{multline}
  L_{\text{\dpce{}}} = \sum_{(u,v)\in E} (2w_{uv}-1) \prod_{l=1}^{\kappa}
  \frac{1 + \tilde x_u^l \tilde x_v^l}{2} \\
  \;+\; \beta v \cdot \frac{1}{\kappa}\sum_{l=1}^{\kappa}
  \left(\frac{1}{n}\sum_i (\tilde x_i^l)^2\right)^{\!2}\!,
  \label{eq:dpce-loss}
\end{multline}
where the product term is a smooth indicator of ``$u,v$ in the same cluster'' (${\approx}1$ when
every bit matches, ${\approx}0$ if any differs), and the coefficient $2w_{uv}-1$ is the key
departure from applying Eq.~\eqref{eq:pce-loss} independently per layer: since $w_{uv}=1-|\rho_{uv}|$
(\S\ref{sec:finance-mapping}), $2w_{uv}-1 = 1-2|\rho_{uv}|$, so strongly correlated pairs
($|\rho|\to1$) get a coefficient $\to -1$, rewarding the optimizer for placing them in the
same cluster, while weakly correlated pairs ($|\rho|\to0$) get coefficient $\to+1$,
rewarding separation. Without this shift, strongly-correlated pairs contribute ${\approx}0$ to
an unshifted loss and scatter randomly across clusters instead of being pulled together. A
node's final label is the integer decoded from its $\kappa$ thresholded bits,
$\text{label}_i = \sum_{l} b_i^l\, 2^{\kappa-1-l}$, and the realized objective is the discrete
weighted multi-way cut, the direct generalization of Eq.~\eqref{eq:cut-value} to more than two
labels.

\begin{figure}[t]
  \centering
  \includegraphics[width=\linewidth]{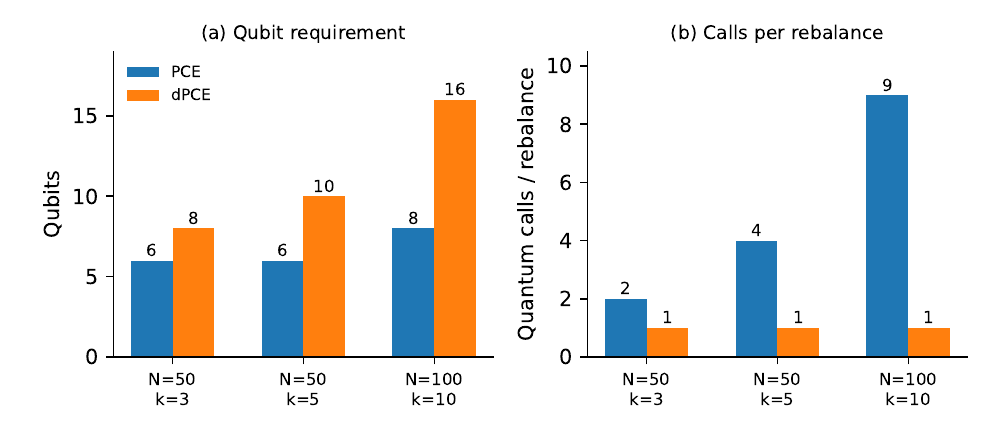}
  \caption{\pce{} vs.\ \dpce{} at \texttt{order}$=2$: (a) qubit requirement and (b) quantum calls
  per rebalance, for the cluster counts used in this paper's smaller-scale experiments. \dpce{}'s
  qubit count is for its one joint call on the $\kappa n$-node layered graph; \pce{}'s is for its
  largest (first) call on the full $n$-node graph. \dpce{} always costs more qubits but exactly 1
  call; \pce{} stays cheaper in qubits but needs $k-1$ calls.}
  \label{fig:qubits-vs-iterations}
\end{figure}

\paragraph{Qubit--iteration trade-off.}
The qubit requirement in Eq.~\eqref{eq:calculate_qubits} is applied to $n$ nodes for \pce{}'s
first split, but to $\kappa n$ nodes for \dpce{}'s only call -- $\dpce{}$ always
needs at least as many qubits as \pce{}'s largest single call, and strictly more whenever
$\kappa>1$ (i.e.\ whenever $k>2$ clusters are requested). Figure~\ref{fig:qubits-vs-iterations}
makes this concrete for the cluster counts used in this paper's smaller-scale experiments.
Against this, \pce{} pays not in qubits but in iterations: producing $k$ clusters costs
\dpce{} exactly 1 quantum call, but \pce{} needs $k-1$ -- at $k=15$ ($N=300$), that is $14$
separate optimization calls per rebalance against \dpce{}'s $1$. Whether this trade-off favors
\pce{} or \dpce{} in wall-clock terms is therefore not obvious from qubit count alone: \dpce{}'s
single call operates on a larger circuit (more qubits, deeper ansatz), while \pce{}'s many calls
each operate on a smaller circuit, with every call after the first acting on a shrinking
sub-cluster (\S\ref{sec:pce-algorithm}). \S\ref{sec:results-scalability} measures this directly
and finds the per-call cost advantage of \pce{}'s smaller circuits dominates the cost of needing
more of them, by close to an order of magnitude at $N=300$ -- i.e.\ the qubit-count gap in
Figure~\ref{fig:qubits-vs-iterations} is a reliable leading indicator of which algorithm will be
cheaper to run, even though it is not, by itself, a runtime measurement.

%% file: sections/03_finance_mapping.tex
\subsection{Market graph construction: assets as nodes, correlation as edges}
Each of the $n$ candidate assets in the investable universe is one node of the graph consumed by
\S\ref{sec:methodology}. Given a trailing window of $T$ daily returns (weighted as in
Eq.~\eqref{eq:window-weight}, introduced below), the edge weight between assets $u$ and $v$ is
\begin{equation}
  w_{uv} = 1 - |\hat\rho_{uv}|,
  \label{eq:edge-weight}
\end{equation}
re-scaled linearly to $[0,1]$ within each window so the graph-cutting landscape has consistent
contrast regardless of the window's absolute correlation level. This sign is what turns the
graph-cutting problem of \S\ref{sec:methodology} into a diversification objective: a
high-weight edge connects a low-correlation (or anti-correlated) pair, so
maximizing the cut -- separating high-weight edges across clusters -- groups together assets
that move together and separates assets that do not. Edges below a minimum $|\hat\rho|$
threshold are dropped entirely rather than given a near-1 weight, so the graph reflects genuine
co-movement, not noise-level correlation treated as a cluster boundary.

\subsection{Streaming rebalancing: windows and step size}
\label{sec:streaming-windows}
\begin{figure}[t]
  \centering
  \includegraphics[width=\linewidth]{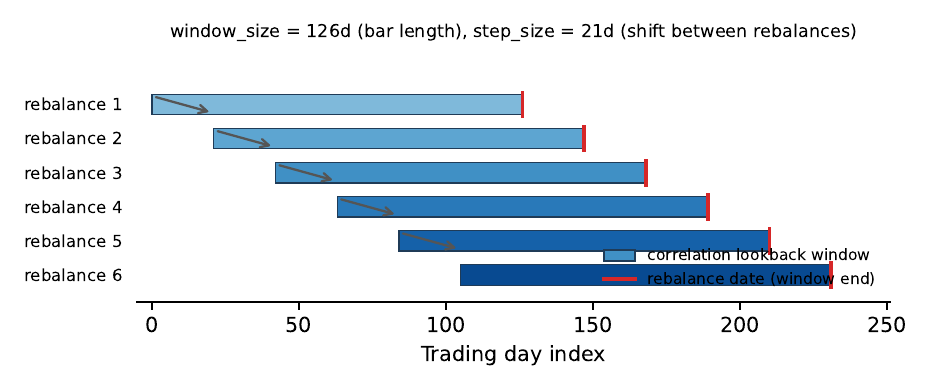}
  \caption{Rolling rebalancing schematic: each bar is one \texttt{window\_size}-day correlation
  lookback window, the red tick its rebalance date; successive windows are offset by
  \texttt{step\_size} days.}
  \label{fig:streaming-window}
\end{figure}

Every result in this paper is produced by a rolling procedure, not a single calculation: at each
rebalance date, the correlation graph above is built from a trailing window of
\texttt{window\_size} daily returns, clustered and converted into a portfolio as described above,
held out-of-sample for \texttt{step\_size} trading days, then the window is advanced by
\texttt{step\_size} and the whole pipeline repeats. Figure~\ref{fig:streaming-window} shows this
mechanically: successive windows overlap by \texttt{window\_size}$-$\texttt{step\_size} days, so
correlation estimates at consecutive rebalances are not independent of each other, and each
window's right edge (red tick) is the date at which a new portfolio is computed and the previous
one is replaced. Unless swept as an independent variable (\S\ref{sec:results}), we use
\texttt{window\_size}$=126$ trading days ($\approx$6 months) and \texttt{step\_size}$=21$
($\approx$1 month), matching common institutional rebalancing cadence. This loop is what
distinguishes every result in this paper from \citet{soloviev2026pce}'s static study: every
rebalance re-issues the full clustering procedure, so solution quality, computational cost, and
the effect of every parameter below accumulate over the life of the backtest rather than being
paid once. Within a window, not every daily observation is treated equally --
\S\ref{sec:method-window} below specifies how the $T=\texttt{window\_size}$ observations are
weighted by recency and traded volume before the correlation matrix of
Eq.~\eqref{eq:edge-weight} is estimated.

\subsection{Empirical properties of the market graph}
\label{sec:market-graph-example}
\begin{figure}[t]
  \centering
  \includegraphics[width=\linewidth]{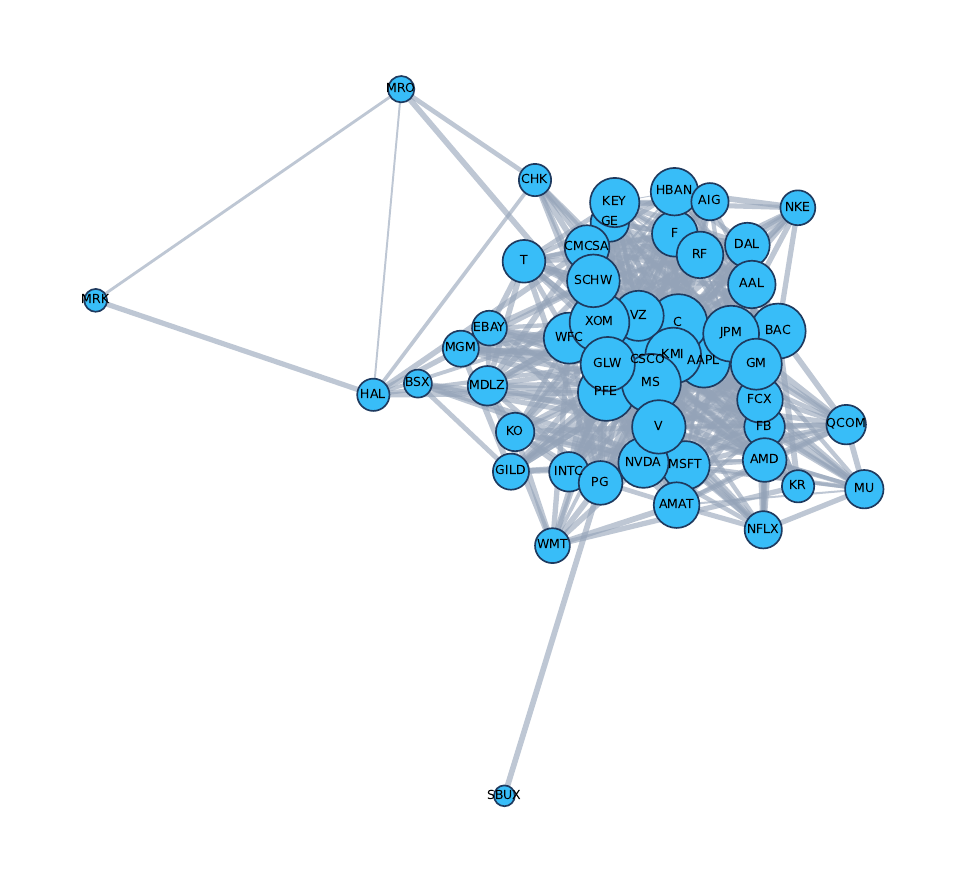}
  \caption{Example market graph, $N=50$ highest-volume S\&P~500 tickers, $126$-day window. Node
  size $\propto$ degree; edge width $\propto$ edge weight $w_{uv}$.}
  \label{fig:market-graph}
\end{figure}

Figure~\ref{fig:market-graph} shows this construction applied to the $N=50$ highest-volume
tickers in our S\&P~500 dataset, over the most recent $126$-day window of the backtest period
(node size scales with degree; edge width scales with weight). Table~\ref{tab:graph-stats}
reports the standard \texttt{networkx} descriptive statistics for this graph. Two features are
worth noting for what they imply about the graph-cutting problem \pce{}/\dpce{} actually solve
at this scale: the graph is dense (density $0.39$, i.e.\ $39\%$ of all possible edges survive
the correlation threshold) and highly clustered (average clustering coefficient $0.70$), with a
single connected component -- almost every large-cap name in this window is correlated with
almost every other one strongly enough to matter, and the handful of outliers (e.g.\ energy and
healthcare names in Figure~\ref{fig:market-graph}'s periphery) are connected to the dense core
through only one or two edges each, rather than being fully isolated. This is a qualitatively
different regime from a sparse, loosely-connected random graph: a dense, highly-clustered graph
with one dominant component is exactly the structure where a small number of cuts can separate
large, cohesive communities, which is what makes a graph-cutting formulation a reasonable
diversification heuristic at this scale, rather than the Pauli encoding's compactness being the
only thing being tested.

\begin{table}[t]
\centering
\caption{\texttt{networkx} descriptive statistics for the graph in Figure~\ref{fig:market-graph}.}
\label{tab:graph-stats}
\small
\begin{tabular}{lr}
\toprule
Nodes & 50 \\
Edges & 483 \\
Density & 0.394 \\
Mean degree & 19.32 \\
Mean clustering coefficient & 0.699 \\
Connected components & 1 (all 50 nodes) \\
Isolated nodes & 0 \\
Mean edge weight $w_{uv}$ & 0.772 \\
\bottomrule
\end{tabular}
\end{table}

\subsection{Portfolio construction from clusters}
The $k$ clusters produced by \pce{} (\S\ref{sec:pce-algorithm}) or \dpce{}
(\S\ref{sec:dpce-algorithm}) are, by the construction above, $k$ internally-correlated,
mutually-diversifying groups of assets. The portfolio for the window is one representative asset
per cluster, equal-weighted, selected by the transaction-cost-aware score of
\S\ref{sec:method-tx} below: the single best-performing name from each diversification bucket,
under a liquidity floor set by the risk-aversion profile of \S\ref{sec:method-risk-aversion}.
This is the portfolio-construction step; it is a classical post-processing of the quantum (or
quantum-simulated) clustering output and does not itself require a quantum call.

\subsection{Transaction costs}
\label{sec:method-tx}
A graph cut determines clusters; it says nothing about the cost of moving from last
window's portfolio to this window's. Ignoring that cost -- re-optimizing asset selection from
scratch every rebalance -- is the implicit assumption of \S\ref{sec:methodology}'s loss
functions, and it is a poor model of real trading: transaction costs (bid-ask spread, market
impact, commissions) are a first-order consideration in any realistic execution
framework~\citep{almgren2000optimal}, and ignoring them in the selection objective systematically
overstates achievable turnover-adjusted returns. We add a stickiness bonus to the selection score
of whichever asset was already held:
\begin{equation}
  \text{score}(a) = \text{Sharpe}_{\text{window}}(a) + \tau \cdot \mathbb{1}[a \in
  \text{held}_{t-1}],
  \label{eq:tx-bonus}
\end{equation}
so a previously-held asset is only displaced when a candidate's in-window Sharpe ratio exceeds
it by more than $\tau$. This is a deliberately simple, single-parameter cost model -- a fixed
Sharpe-ratio-denominated hurdle rather than an explicit price-impact function -- chosen because
it composes directly with the cluster-selection score already in use, rather than requiring a
separate execution-cost model layered on afterward; \S\ref{sec:results-tx} evaluates its effect
on realized turnover directly.

Two metrics used throughout this score and the rest of the paper are worth making precise.
$\text{Sharpe}_{\text{window}}(a)$ above, and every portfolio-level Sharpe ratio reported in
\S\ref{sec:results}, use the same convention: annualized mean return divided by annualized
volatility of the relevant daily return series ($\sqrt{252}$ scaling), with an implicit
risk-free rate of zero. This inflates reported Sharpes somewhat relative to a true excess-return
Sharpe, particularly during the $N=5{,}000$ backtest's 2021--2024 window when risk-free rates
were not negligible, but the same convention is applied uniformly to every strategy and
baseline, so relative comparisons between them are unaffected. Turnover, reported throughout
\S\ref{sec:results} and Appendix~\ref{app:data}, is the fraction of the $n_{\text{clusters}}$
held positions replaced at a given rebalance relative to the previous one.

\subsection{Risk-aversion profile}
\label{sec:method-risk-aversion}
Classical portfolio theory represents an investor's risk preference with a scalar risk-aversion
coefficient $\lambda$ in a mean-variance utility $U = \mathbb{E}[r] - \tfrac{\lambda}{2}
\text{Var}[r]$~\citep{markowitz1952portfolio}: larger $\lambda$ trades expected return for lower
variance. Our implementation exposes a single scalar $\rho\in[0,1]$ in the same spirit -- a
user-facing dial between an aggressive and a conservative profile -- but $\rho$ is an
operational proxy, not Markowitz's $\lambda$ itself: rather than entering a
mean-variance objective directly, $\rho$ drives three mechanisms in the graph-construction and
selection pipeline simultaneously: the recency/volume weighting of the correlation estimate
(\S\ref{sec:method-window}), a liquidity floor excluding the bottom $25\rho\,\%$ of candidates by
traded volume, and the transaction-cost bonus itself, $\tau = \tau_{\max}\cdot\rho$ (default
$\tau_{\max}=0.05$, overridable independently of $\rho$). Coupling three mechanisms to one dial
is a deliberate simplicity choice for a user-facing
control, trading the interpretability of a single textbook risk-aversion coefficient for an
operational control that is easier to wire into a real system; \S\ref{sec:results} tests whether
the result actually behaves as a smooth control or only a coarse one, and compares it against
decoupling the mechanisms.

\subsection{Correlation-window weighting}
\label{sec:method-window}
\begin{figure}[t]
  \centering
  \includegraphics[width=\linewidth]{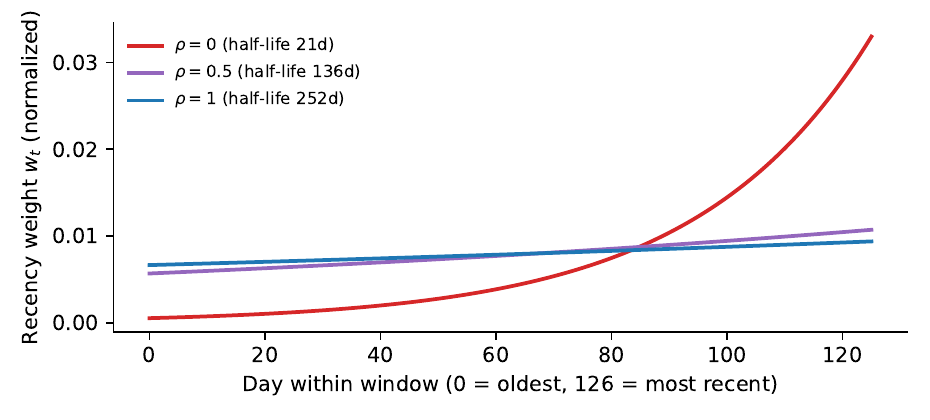}
  \caption{Recency component of the within-window observation weight $w_t$
  (Eq.~\eqref{eq:window-weight}), normalized, across a $T=126$-day window, at three
  risk-aversion levels $\rho$.}
  \label{fig:window-weighting}
\end{figure}

Rather than an unweighted Pearson correlation over the lookback window, each of the $T$
observations in the window is weighted by recency and relative traded volume -- an exponentially
weighted moving estimator in the tradition of the RiskMetrics
methodology~\citep{riskmetrics1996}, which popularized exponential decay weighting for
volatility and covariance estimation in risk management precisely because it down-weights stale
observations smoothly rather than via a hard window cutoff:
\begin{align}
  \lambda &= 2^{-1/\tau(\rho)}, \qquad \tau(\rho) = 21 + 231\,\rho \;\;\text{(days)}, \\
  w_t &\;\propto\; \lambda^{\,T-1-t} \cdot \Big(\tfrac{\text{vol}_t}{\overline{\text{vol}}_t}\Big)^{\alpha(\rho)},
  \qquad \alpha(\rho) = 2.0 - 2.5\,\rho,
  \label{eq:window-weight}
\end{align}
normalized to $\sum_t w_t = 1$. $\tau(\rho)$ is the recency half-life: $\rho=0$ gives a 21-day
half-life (near-term observations dominate, in the spirit of a fast-adapting EWMA estimator),
$\rho=1$ gives 252 days (close to equal-weighting the full window, the conservative limit).
Figure~\ref{fig:window-weighting} shows the resulting recency-weight profile within a single
$T=126$-day window at three representative $\rho$ values: at $\rho=0$ the most recent days
dominate the correlation estimate sharply, while at $\rho=0.5$ and $\rho=1$ the profile is
nearly flat, close to the unweighted average implicit in a plain Pearson correlation.
$\alpha(\rho)$ ranges from $+2.0$ (strongly upweight high-volume days) to $-0.5$ (downweight
them) -- the negative sign at high $\rho$ is a deliberate departure from standard EWMA volume
weighting: volume spikes coincide with panic/crisis periods where realized correlations
spuriously converge toward $\pm1$ (the well-documented phenomenon of correlation breakdown
during market stress), so a conservative profile should discount those days rather than
emphasize them. The weighted correlation matrix is then $\hat\rho_{uv} = \sum_t w_t\,(r_u(t)-\bar
r_u)(r_v(t)-\bar r_v) / (\hat\sigma_u \hat\sigma_v)$, with $\bar r$ and $\hat\sigma$ computed
under the same weights.

\subsection{Classical optimizer and ansatz}
\label{sec:method-decisions}
We use \texttt{COBYQA} as the classical outer-loop
optimizer throughout (gradient-free, appropriate for the noisy, shot-based objective of
\S\ref{sec:results-shots}, and for an optimization loop that in a streaming deployment must run
unattended at every rebalance without per-instance tuning); a hardware-efficient
\texttt{LayeredHEA} ansatz as the default parameterized circuit (3 repetition layers unless
stated otherwise), with a fixed-depth \texttt{BrickworkPCE} alternative available but not the
default; and warm-starting (reusing the previous rebalance's converged parameters as the next
window's initialization) enabled by default, whose effect on solution quality and convergence
speed is measured directly in \S\ref{sec:results-warmstart} rather than assumed.

Table~\ref{tab:ansatz-comparison} quantifies the ansatz choice directly, computed from the
circuit implementations themselves rather than estimated: at every qubit count tested (6, 8, and
15 qubits, matching $(N,\text{order})$ configurations used elsewhere in this paper),
\texttt{LayeredHEA} uses substantially fewer parameters and total gates than
\texttt{BrickworkPCE}. \texttt{BrickworkPCE}'s depth stays fixed at 18 regardless of qubit count
-- it has no repetition-layer parameter; its structure is three fixed single-qubit rotation
layers interleaved with brickwork entangling sublayers -- while \texttt{LayeredHEA}'s depth grows
with $n$, because its circular linear entangler is a chain of $n$ sequential two-qubit gates per
repetition layer. We also compute two standard circuit-quality diagnostics (200--300
random-parameter statevector samples per configuration, following the methodology
of~\citet{sim2019expressibility}): the Meyer--Wallach entangling capability $Q\in[0,1]$ (higher
means more multi-qubit entanglement on average over random parameters) and the expressibility KL
divergence to a Haar-random ensemble (lower means the circuit's output-state distribution is
closer to Haar-random, i.e.\ more expressive). \texttt{LayeredHEA} reaches a substantially higher
entangling capability than \texttt{BrickworkPCE} at every qubit count ($Q\approx0.91$--$0.95$
vs.\ $0.58$--$0.61$), despite its lower gate count -- a consequence of the circular entangling
topology coupling every qubit to its neighbors every layer, rather than the brickwork pattern's
local, alternating-pair structure. Expressibility does not show the same one-sided picture: at
$n=6$ qubits \texttt{BrickworkPCE} is marginally closer to Haar (KL $0.107$ vs.\ $0.245$), the
ordering reverses by $n=8$ (\texttt{LayeredHEA} at $0.045$ vs.\ $0.103$), and at $n=15$ both are
statistically indistinguishable from Haar at the sample sizes used here (KL $\approx0$ for both)
-- finite-sample KL estimation saturates once the $2^n$-dimensional Hilbert space is large enough
that randomly drawn fidelity pairs are essentially always near zero, a known limitation of this
diagnostic at larger qubit counts rather than evidence the two ansätze are equally expressive in
any exact sense. Taken together, the entangling-capability gap is the more decisive and
consistent finding of the two, and is consistent with \texttt{LayeredHEA} being the default
throughout this paper.

\begin{table}[t]
\centering
\caption{Ansatz resource counts and circuit-quality diagnostics, by qubit count $n$ (3 repetition
layers for \texttt{LayeredHEA}; entangling capability and expressibility computed from 200--300
random-parameter statevector samples each).}
\label{tab:ansatz-comparison}
\small
\resizebox{\linewidth}{!}{%
\begin{tabular}{llrrrrr}
\toprule
Ansatz & $n$ & Params & Gates (1q/2q) & Depth & Ent.\ cap.\ $Q$ & Expr.\ KL \\
\midrule
LayeredHEA    & 6  & 18 & 18/18  & 21 & 0.906 & 0.245 \\
BrickworkPCE  & 6  & 26 & 58/16  & 18 & 0.579 & 0.107 \\
LayeredHEA    & 8  & 24 & 24/24  & 27 & 0.933 & 0.045 \\
BrickworkPCE  & 8  & 35 & 79/22  & 18 & 0.599 & 0.103 \\
LayeredHEA    & 15 & 45 & 45/45  & 48 & 0.945 & $\sim$0 \\
BrickworkPCE  & 15 & 66 & 150/42 & 18 & 0.614 & $\sim$0 \\
\bottomrule
\end{tabular}}
\end{table}

\subsection{Baseline strategy}
Throughout, the comparison baseline is equal-weight Buy~\&~Hold over the same candidate
universe, held fixed for the full backtest -- the realistic alternative to running any
clustering procedure at all, static or streaming.

\subsection{Data sources}
\label{sec:data}
Experiments use real daily-OHLCV equity data from three sources, described alongside the results
that use them (\S\ref{sec:results}): an S\&P~500, 5-year dataset (2013-02-08--2018-02-07) for
$N\leq300$, the market-graph example above, and the methodology-parameter studies; a
2{,}070-ticker, 1980--2022 dataset for the $N=1{,}601$ large-scale result; and a broader
9{,}315-ticker, 1962--2024 dataset for the $N=5{,}000$ result
(\S\ref{sec:results-scalability}), from which the 2021--2024 backtest window is drawn. In every
case only tickers with complete (gap-free) price history over the specific backtest window are
retained, with an additional filter dropping any ticker showing an implausible single-day move
($>50\%$, even in split/dividend-adjusted prices) as a data-quality check. Appendix~\ref{app:data}
reports descriptive statistics -- including pairwise return-correlation distributions -- for all
three datasets.

%% file: sections/04_results.tex
Unless stated otherwise, results use the S\&P~500 5-year dataset (2013-02-08--2018-02-07,
\S\ref{sec:data}), \texttt{window\_size}$=126$d, \texttt{step\_size}$=21$d, and
$n_{\text{clusters}}$ scaling with $N$ as $\{3,5,10,15\}$ for $N\in\{50,100,200,300\}$.

\subsection{Window-length sensitivity of streaming \pce{}}
\label{sec:results-pce-streaming}
\begin{figure}[t]
  \centering
  \includegraphics[width=\linewidth]{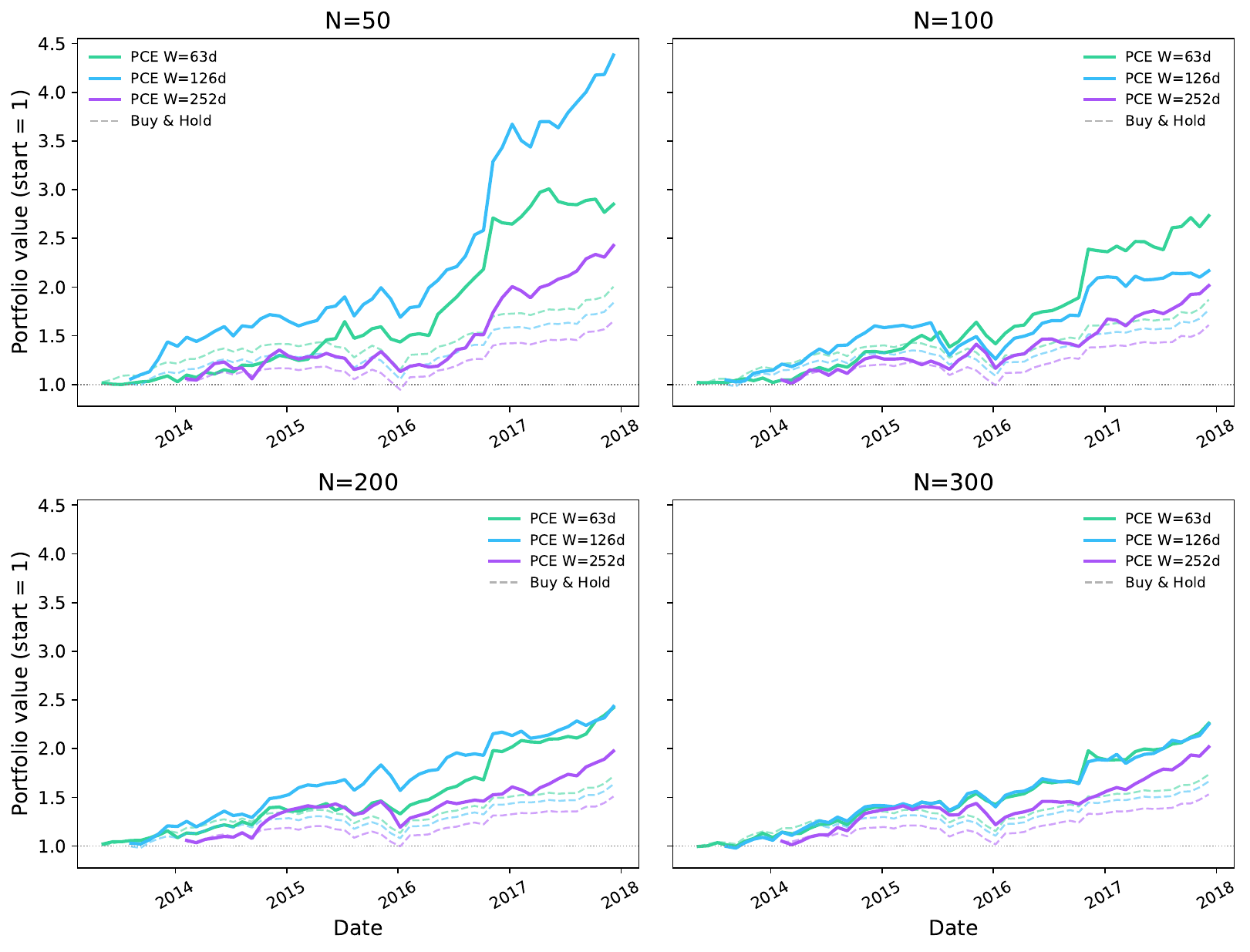}
  \caption{Streaming \pce{}, cumulative return vs.\ Buy \& Hold, by universe size $N$ and
  correlation window length.}
  \label{fig:pce-windows}
\end{figure}

Figure~\ref{fig:pce-windows} shows streaming \pce{} against Buy~\&~Hold across universe size
$N\in\{50,100,200,300\}$ and correlation window $\in\{63,126,252\}$d. \pce{} beats
Buy~\&~Hold at every $(N, \text{window})$ combination tested. The $126$-day window is the
consistent best performer across all four universe sizes; the shortest window ($63$d) is
competitive at small $N$ but falls behind as $N$ grows, and the longest window ($252$d) is the
weakest throughout.

\begin{figure}[t]
  \centering
  \includegraphics[width=\linewidth]{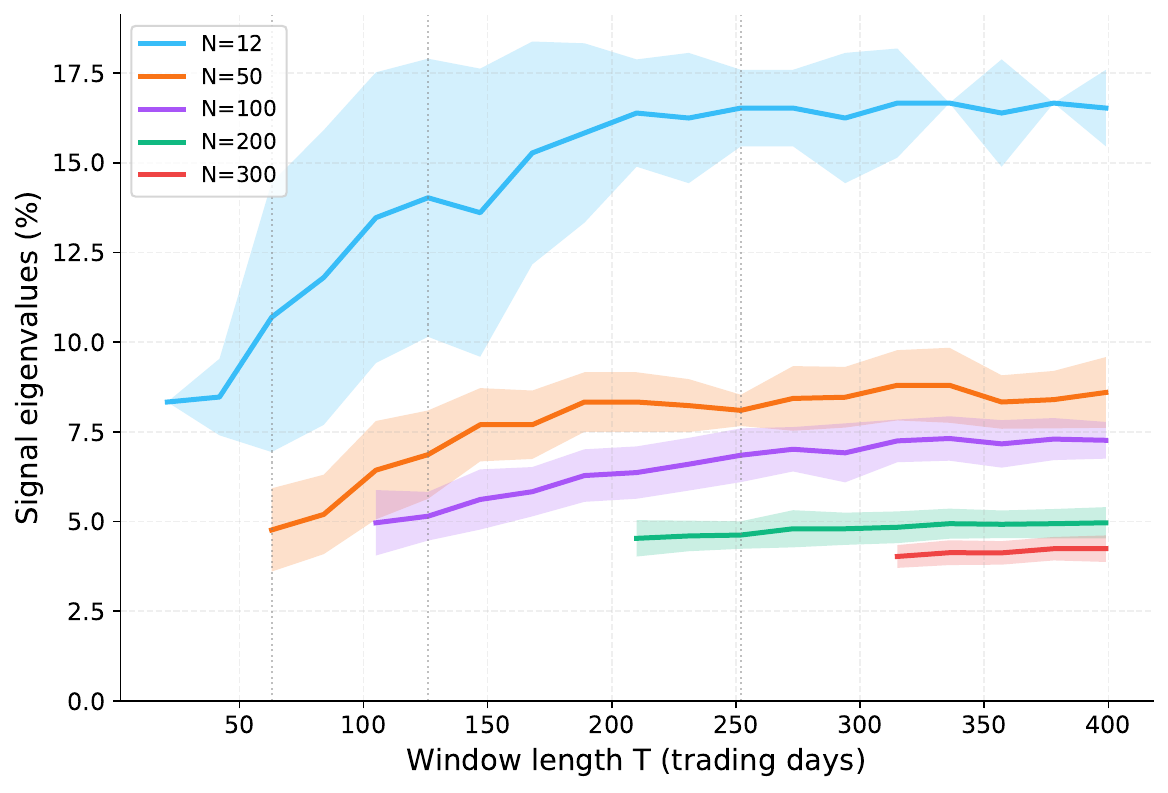}
  \caption{Fraction of sample correlation-matrix eigenvalues above the Marchenko--Pastur noise
  threshold $\lambda^+$, vs.\ window length $T$, by universe size $N$ (mean $\pm$ 1 std across
  60 day-level bootstrap resamples). Lines start where $T$ first exceeds $N$.}
  \label{fig:rmt-signal}
\end{figure}

A natural first hypothesis is that longer windows simply give a statistically cleaner
correlation estimate: more observations per asset should push more eigenvalues of the sample
correlation matrix above the Marchenko--Pastur noise threshold $\lambda^+=(1+\sqrt{q})^2$,
$q=N/T$, separating genuine co-movement structure from sampling noise. Figure~\ref{fig:rmt-signal}
tests this directly: the fraction of eigenvalues classified as signal does increase monotonically
with window length $T$ at every $N$ tested, but it plateaus by $T\approx150$--$250$ days and is
essentially flat beyond that -- it does not predict that the longest window ($252$d) should
perform worst, which is exactly what Figure~\ref{fig:pce-windows} shows. This diagnostic is also
only well-posed at $N\leq300$: it requires $T$ strictly greater than $N$ to even compute a
full-rank sample correlation matrix, so the two largest universe sizes used in
\S\ref{sec:results-scalability} ($N=1{,}601$ and $N=5{,}000$) fall outside where it applies at
all. Taken together, the empirical window-length result is better explained as a
staleness-versus-noise trade-off than a pure noise-reduction effect: a longer window does yield a
less noisy correlation estimate by this measure, but it also smooths over more recent regime
change, and the recency-weighting mechanism of \S\ref{sec:method-window}
(Figure~\ref{fig:window-weighting}) exists precisely to recover some of that responsiveness
without shortening the window itself. We do not have a clean decomposition of the $126$d optimum
into these two competing effects and flag it as an open question.

\subsection{Risk-aversion sensitivity}
\label{sec:results-risk-aversion}
\begin{figure}[t]
  \centering
  \includegraphics[width=\linewidth]{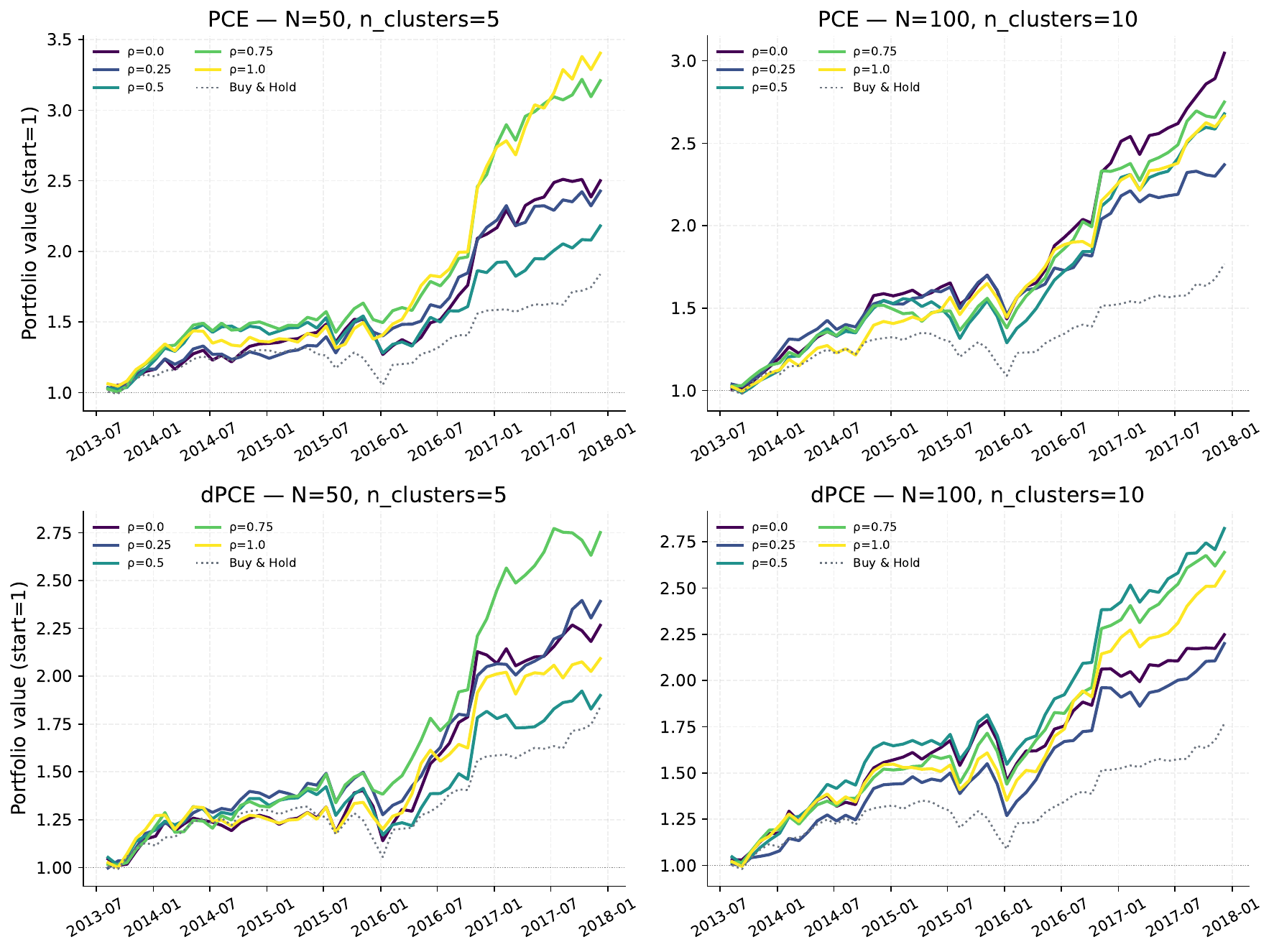}
  \caption{Cumulative return by risk-aversion level $\rho$, \pce{} and \dpce{}. Every curve
  beats Buy \& Hold (dotted) regardless of $\rho$.}
  \label{fig:risk-aversion}
\end{figure}

Sweeping the risk-aversion profile $\rho\in\{0,0.25,0.5,0.75,1.0\}$
(\S\ref{sec:method-risk-aversion}) at $N\in\{50,100\}$, every configuration of both algorithms
beats Buy~\&~Hold regardless of $\rho$ (Figure~\ref{fig:risk-aversion}) -- $\rho$ is not a
return-destroying safety control, it is a profile selector among already-favorable outcomes.
Table~\ref{tab:risk-aversion-sharpe} reports the corresponding Sharpe ratios: they do not order
monotonically in $\rho$ for either algorithm (e.g.\ \pce{}, $N=50$: $6.30\to6.93\to5.42\to7.42\to7.77$),
the one partial exception being that the $\rho=1$ endpoint is the single best or
near-best Sharpe in three of the four rows shown. We return to exactly how coarse this control
is, and what a better one looks like, in \S\ref{sec:results-rebalance-vs-risk-aversion}.

\begin{table}[t]
\centering
\caption{Sharpe ratio by risk-aversion level $\rho$. Buy \& Hold is $\rho$-invariant (reported
once per row for reference).}
\label{tab:risk-aversion-sharpe}
\small
\resizebox{\linewidth}{!}{%
\begin{tabular}{llrrrrrr}
\toprule
Algo & $N$ & $\rho{=}0$ & $\rho{=}.25$ & $\rho{=}.5$ & $\rho{=}.75$ & $\rho{=}1$ & B\&H \\
\midrule
PCE  & 50  & 6.30 & 6.93 & 5.42 & 7.42 & 7.77 & 5.09 \\
dPCE & 50  & 5.24 & 7.13 & 4.36 & 7.40 & 5.04 & 5.09 \\
PCE  & 100 & 8.41 & 7.26 & 7.30 & 7.78 & 7.89 & 5.17 \\
dPCE & 100 & 6.17 & 6.32 & 8.22 & 7.60 & 7.16 & 5.17 \\
\bottomrule
\end{tabular}}
\end{table}

\subsection{Transaction-cost sensitivity}
\label{sec:results-tx}
Trading is not free: every change in holdings pays the bid-ask spread, market impact, and
commissions, and this real-world cost scales directly with how much of the portfolio is replaced
at each rebalance~\citep{almgren2000optimal} -- this is precisely what motivates the stickiness
bonus of Eq.~\eqref{eq:tx-bonus} in the first place. Whether suppressing turnover also improves
\emph{realized} backtest return is not obvious a priori: the backtests reported throughout this
paper are frictionless (no trading cost is actually deducted from the simulated return), so
adding the bonus can only change the outcome by changing which marginal asset gets selected each
rebalance, not by avoiding a cost that was never charged in the first place. Consistent with
that, total return moves with no consistent sign when the bonus is added ($-53$ to $+56$
percentage points across the 10 configurations below) -- this experiment does not show, and is
not designed to show, a clean return benefit.

What it does show, robustly and in every configuration tested, is the reduction in turnover
itself (Table~\ref{tab:turnover-reduction}; mean $11.3\%\pm2.8\%$ across 10 configurations
spanning both algorithms, $N\in\{50,100\}$, and multiple $n_{\text{clusters}}$): the stickiness
bonus reliably makes the strategy trade less. Since real transaction costs are charged against
exactly this quantity -- assets bought and sold, not simulated gross return -- this turnover
reduction is the relevant result for a live deployment, independent of whether it is visible in
the frictionless backtest return above.

\begin{table}[t]
\centering
\caption{Mean turnover with vs.\ without transaction-cost-aware selection.}
\label{tab:turnover-reduction}
\small
\resizebox{\linewidth}{!}{%
\begin{tabular}{llrrrr}
\toprule
Algo & $N$ & $n_{\text{cl}}$ & Turnover (w/o) & Turnover (w/) & Reduction \\
\midrule
PCE  & 50  & 5  & 0.525 & 0.468 & 10.8\% \\
PCE  & 50  & 10 & 0.496 & 0.419 & 15.6\% \\
PCE  & 100 & 10 & 0.570 & 0.513 & 9.9\%  \\
PCE  & 100 & 15 & 0.521 & 0.469 & 9.9\%  \\
PCE  & 100 & 20 & 0.482 & 0.441 & 8.6\%  \\
dPCE & 50  & 5  & 0.543 & 0.475 & 12.5\% \\
dPCE & 50  & 10 & 0.500 & 0.440 & 12.1\% \\
dPCE & 100 & 10 & 0.558 & 0.472 & 15.5\% \\
dPCE & 100 & 15 & 0.553 & 0.492 & 11.1\% \\
dPCE & 100 & 20 & 0.516 & 0.481 & 6.8\%  \\
\midrule
\textbf{All configs} & -- & -- & \textbf{0.526} & \textbf{0.467} & \textbf{11.3\%$\pm$2.8\%} \\
\bottomrule
\end{tabular}}
\end{table}

\subsection{Window-length sensitivity of streaming \dpce{}}
\label{sec:results-dpce-streaming}
\begin{figure}[t]
  \centering
  \includegraphics[width=\linewidth]{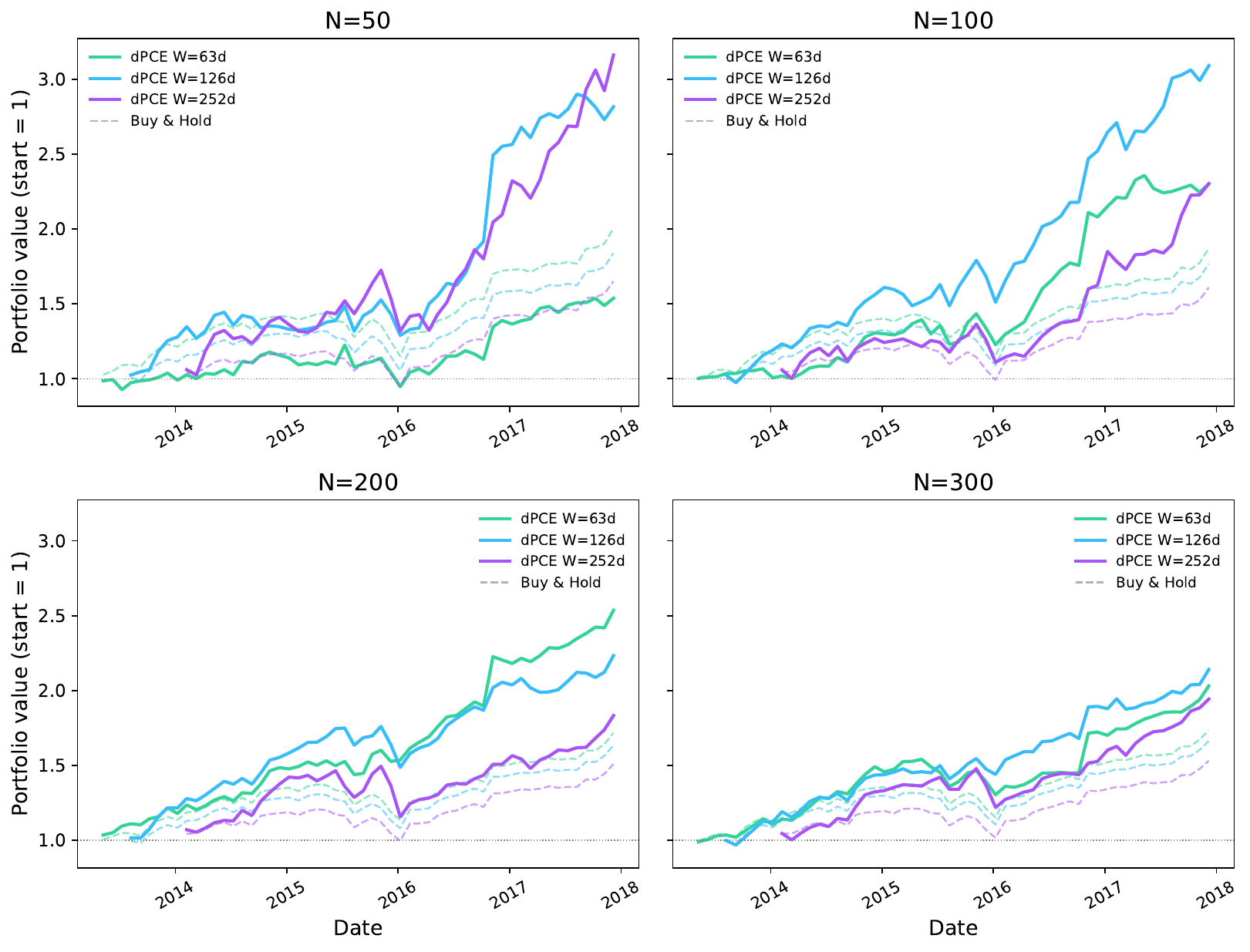}
  \caption{Streaming \dpce{}, cumulative return vs.\ Buy \& Hold, by universe size $N$ and
  correlation window length.}
  \label{fig:dpce-windows}
\end{figure}

Repeating \S\ref{sec:results-pce-streaming} with \dpce{} (Figure~\ref{fig:dpce-windows}) gives
the same qualitative picture: \dpce{} beats Buy~\&~Hold at every tested $(N,\text{window})$, and
$126$d is again the best window at every $N$. \dpce{} reaches higher peak returns than \pce{} at
small $N$ (e.g.\ $N=50$, $W{=}126$d: \dpce{} ${\sim}3.2\times$ vs.\ \pce{} ${\sim}4.4\times$ --
comparable, trading the lead window-to-window), but, as \S\ref{sec:results-scalability} shows,
at a computational cost that scales far worse with $N$.

\subsection{\pce{} vs.\ \dpce{}: direct comparison}
\label{sec:results-comparison}
\begin{figure}[t]
  \centering
  \includegraphics[width=\linewidth]{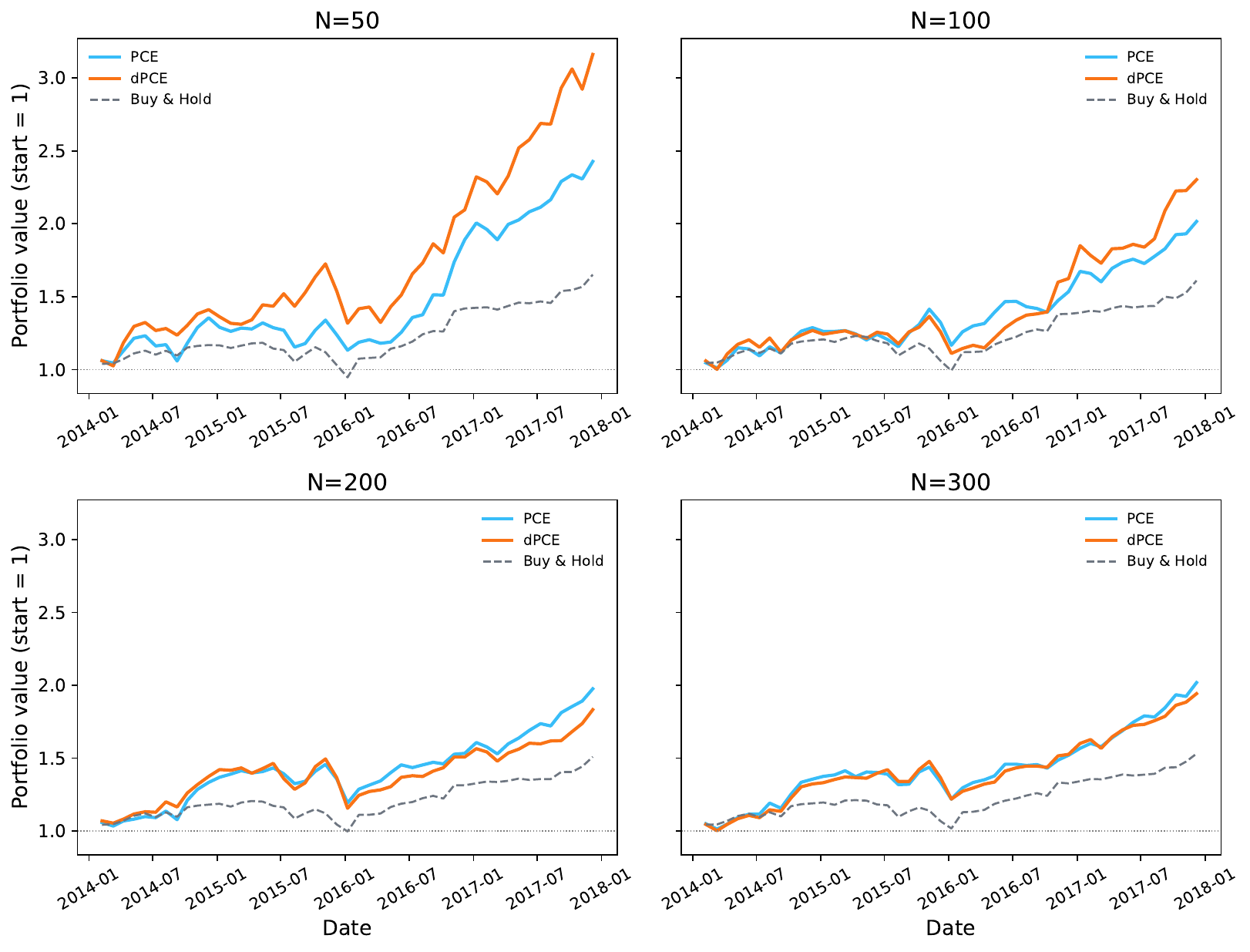}
  \caption{\pce{} vs.\ \dpce{} vs.\ Buy \& Hold, $W=126$d, by universe size $N$.}
  \label{fig:pce-vs-dpce}
\end{figure}

Figure~\ref{fig:pce-vs-dpce} places both algorithms on the same axes at their shared best window
($126$d). The two track each other closely at every $N$, trading the lead rather than one
uniformly dominating -- \dpce{} ahead for most of the backtest at $N=50$, the two nearly
indistinguishable by $N=300$. Both remain comfortably above Buy~\&~Hold throughout this $N$
range. Section~\ref{sec:results-scalability} is the better instrument for separating the two
algorithms: it shows this near-parity in return coexists with an order-of-magnitude
difference in runtime scaling, and that \dpce{} recovers more Sharpe from frequent
rebalancing than \pce{} does.

\subsection{Rebalance cadence vs.\ risk-aversion as a control}
\label{sec:results-rebalance-vs-risk-aversion}
\S\ref{sec:results-risk-aversion} showed $\rho$ gives non-monotonic Sharpe control. Testing
rebalance cadence (\texttt{step\_size}) as an alternative control instead: turnover increases
monotonically with step size (rank correlation $1.00$ at a representative configuration,
holding across a grid of configurations), and total return is negatively rank-correlated with
step size in every configuration tested -- infrequent rebalancing is a dominated strategy
(worse return and higher turnover), never the best-performing choice at any tested
configuration. This makes \texttt{step\_size} the cleaner, more predictable control of the two:
where $\rho$ moves three coupled mechanisms at once and lands non-monotonically on Sharpe,
\texttt{step\_size} moves one and lands monotonically on both turnover and return.

\begin{figure}[t]
  \centering
  \includegraphics[width=\linewidth]{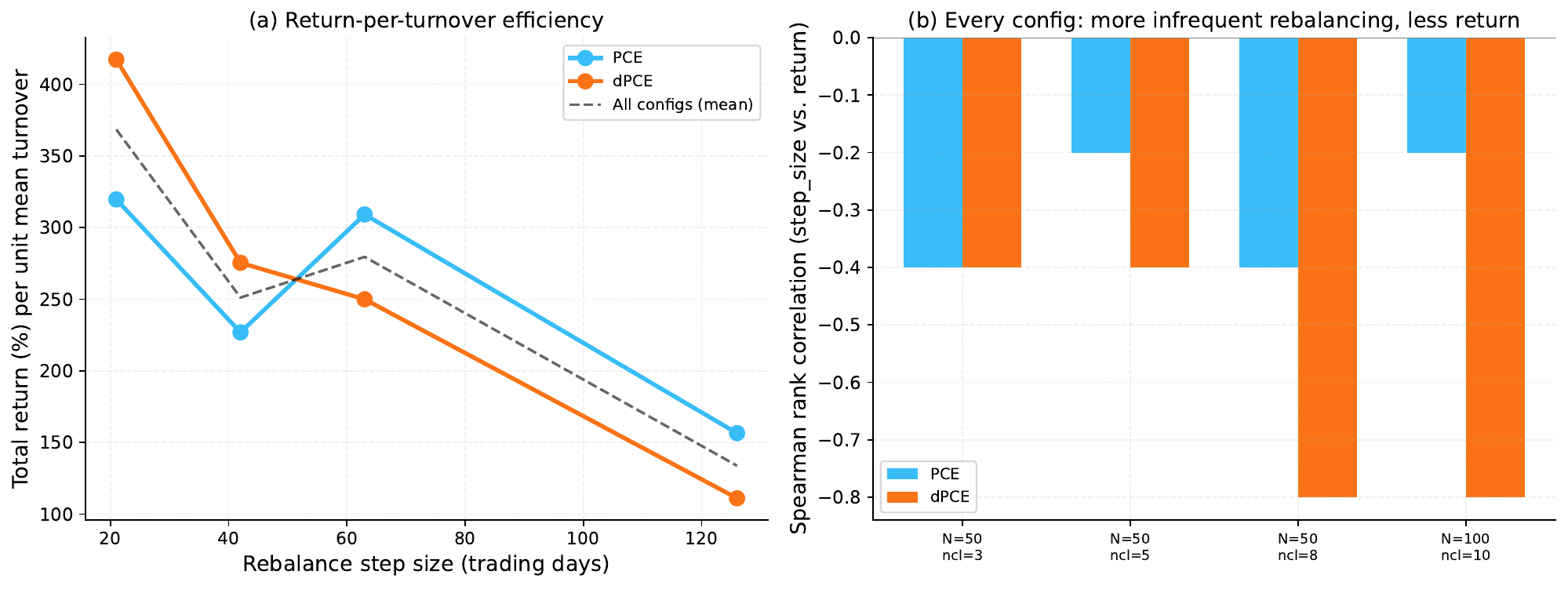}
  \caption{(a) Total return per unit mean turnover vs.\ rebalance step size, \pce{} and \dpce{}
  (dashed: mean across all configurations). (b) Spearman rank correlation between step size and
  total return, by configuration -- every bar is negative, and \dpce{}'s are consistently more
  negative than \pce{}'s.}
  \label{fig:stepsize-efficiency}
\end{figure}

Figure~\ref{fig:stepsize-efficiency}a shows this is a directional effect, not a strictly
monotonic one: return per unit of turnover is highest at the shortest step size (21d) and lowest
at the longest (126d) for both algorithms, but with a local rebound at 63d rather than a smooth
decline throughout. Figure~\ref{fig:stepsize-efficiency}b shows the two algorithms do not share
this sensitivity equally: \pce{}'s rank correlation between step size and return stays in a
narrow $-0.2$ to $-0.4$ band across every tested configuration, while \dpce{}'s ranges from
$-0.4$ to $-0.8$ and grows more negative as $N$ and $n_{\text{clusters}}$ increase (mean
$\approx-0.6$ vs.\ \pce{}'s $\approx-0.3$, roughly twice as sensitive) -- the same asymmetry
visible qualitatively in Figure~\ref{fig:pce-vs-dpce}. Practically, this reframes
\S\ref{sec:results-comparison}'s near-parity between the two algorithms as conditional on
rebalancing frequently: \dpce{} needs a short \texttt{step\_size} to realize its performance,
while \pce{} degrades more gracefully under a coarser cadence, which argues for \pce{} as the
safer default whenever operational or computational constraints force infrequent rebalancing.

\subsection{Warm-start initialization}
\label{sec:results-warmstart}
\begin{figure}[t]
  \centering
  \includegraphics[width=\linewidth]{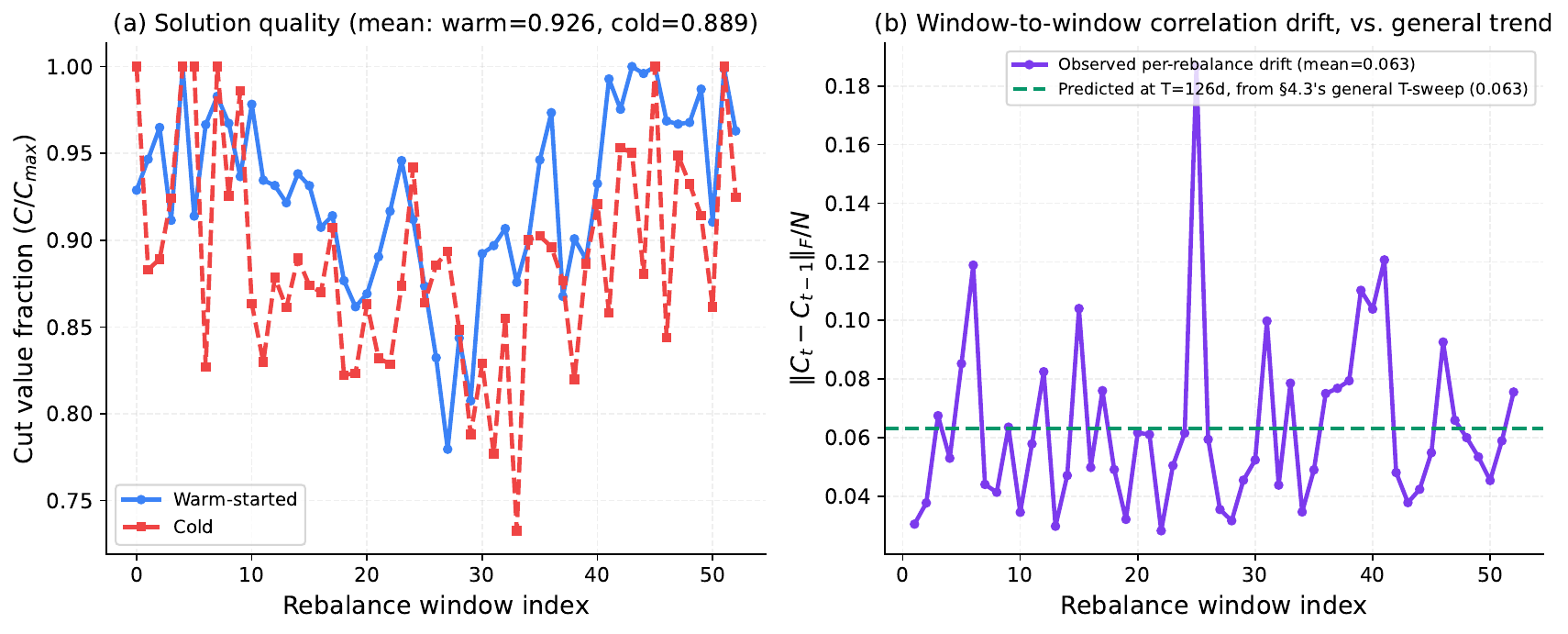}
  \caption{(a) Warm-started vs.\ cold cut-value quality, by rebalance window. (b) Observed
  correlation drift between consecutive windows vs.\ the general drift-vs-window-length trend.}
  \label{fig:warm-start}
\end{figure}

Reusing the previous window's converged ansatz parameters as the next window's initialization
(``warm-starting'') raises mean cut-value quality ($0.926$ vs.\ $0.889$ cold) but does not
reduce the number of optimizer iterations to convergence -- both converge, or hit the
iteration cap, at an identical count every window (Figure~\ref{fig:warm-start}a). This is a
genuine, honest mixed result: warm-starting is worth keeping for quality, but it is not the
convergence-speed win it is often assumed to be. Figure~\ref{fig:warm-start}b contextualizes why
warm-starting is plausible at all: the Frobenius distance between consecutive windows'
correlation matrices, $\lVert C_t - C_{t-1}\rVert_F / N$, is small and stable
(mean $0.063$) -- consecutive windows do not look very different from each other, which is the
precondition for the previous solution being a good initialization for the next one, even
though that precondition alone is not sufficient to speed up convergence here.

\subsection{Shot-count sensitivity}
\label{sec:results-shots}
\begin{figure}[t]
  \centering
  \includegraphics[width=\linewidth]{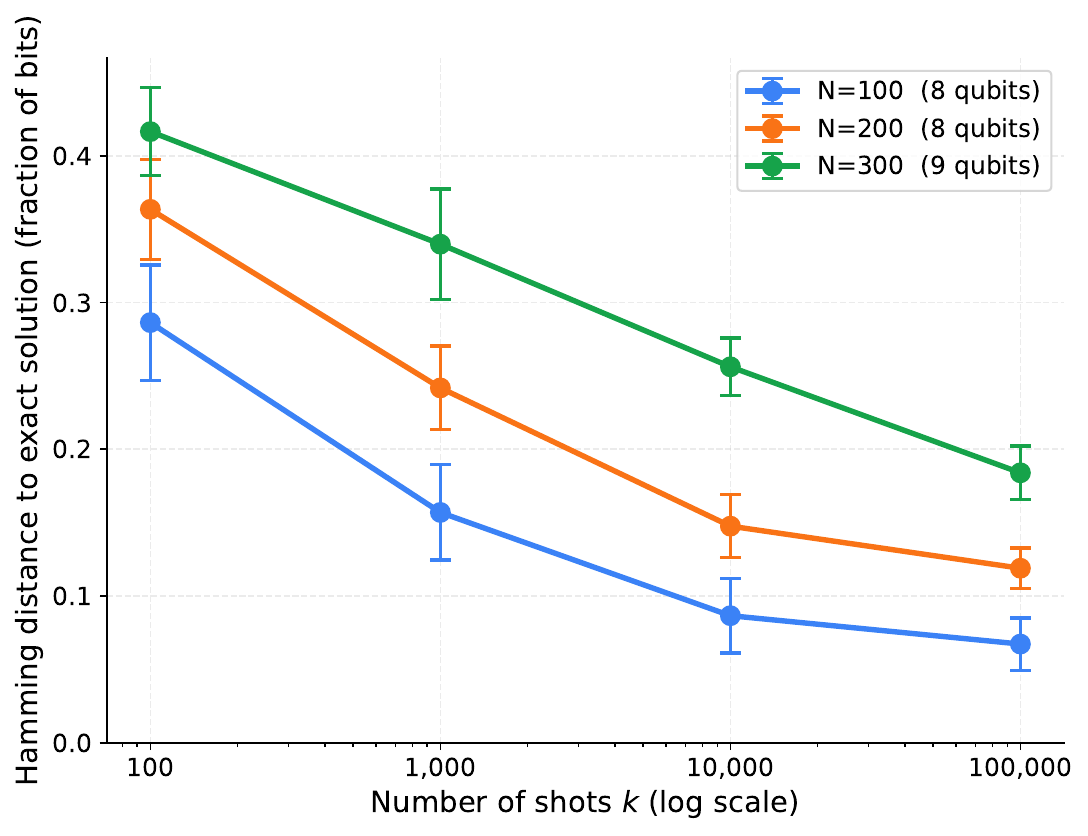}
  \caption{Hamming distance to the exact solution vs.\ shot count, mean $\pm$ std, by $N$.}
  \label{fig:hamming}
\end{figure}

Following \citet{sciorilli2024pce}'s own methodology -- train with exact, noiseless simulation,
apply shots only at final readout -- we train \pce{} to convergence with a noiseless simulator
at $N\in\{100,200,300\}$, freeze the optimized parameters, then re-measure the identical circuit
with $k\in\{100,1{,}000,10{,}000,100{,}000\}$ shots, 30 trials per $(N,k)$. Hamming distance
between the shot-recovered bitstring and the exact solution shrinks monotonically through
$10^5$ shots at every $N$ (Figure~\ref{fig:hamming}) -- more shots reliably recover the exact
bitstring more faithfully, with no sign of saturation even at the highest shot count tested.

\begin{figure}[t]
  \centering
  \includegraphics[width=\linewidth]{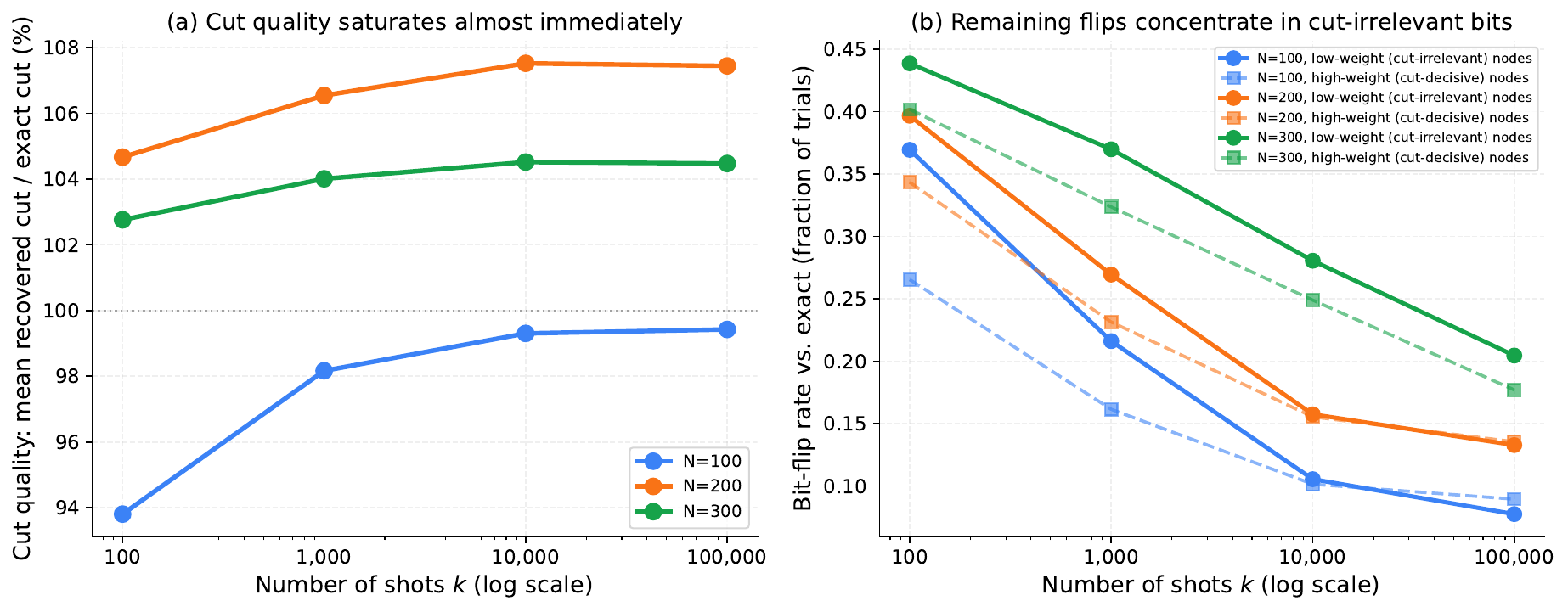}
  \caption{(a) Cut-value quality (as a fraction of the exact value) saturates by $\sim$1{,}000
  shots. (b) The remaining bit-flips at high shot counts concentrate on cut-irrelevant nodes.}
  \label{fig:cut-quality}
\end{figure}

Whether good solutions can be guaranteed with fewer shots is not answered by the
bitstring-recovery metric above: the portfolio-relevant cut-value quality tells a different story
(Figure~\ref{fig:cut-quality}). Mean recovered cut reaches $94$--$107\%$ of the exact value by
just $1{,}000$ shots at every $N$, and stays flat out to $10^5$. The mechanism
(Figure~\ref{fig:cut-quality}b) is that the bits still flipping at high shot counts
disproportionately belong to low-edge-weight (diversification-irrelevant) nodes, whose
expectation value was already near zero -- flipping them barely changes which assets end up in
which cluster. Hamming distance treats every bit equally; the portfolio objective does not.

\begin{figure}[t]
  \centering
  \includegraphics[width=\linewidth]{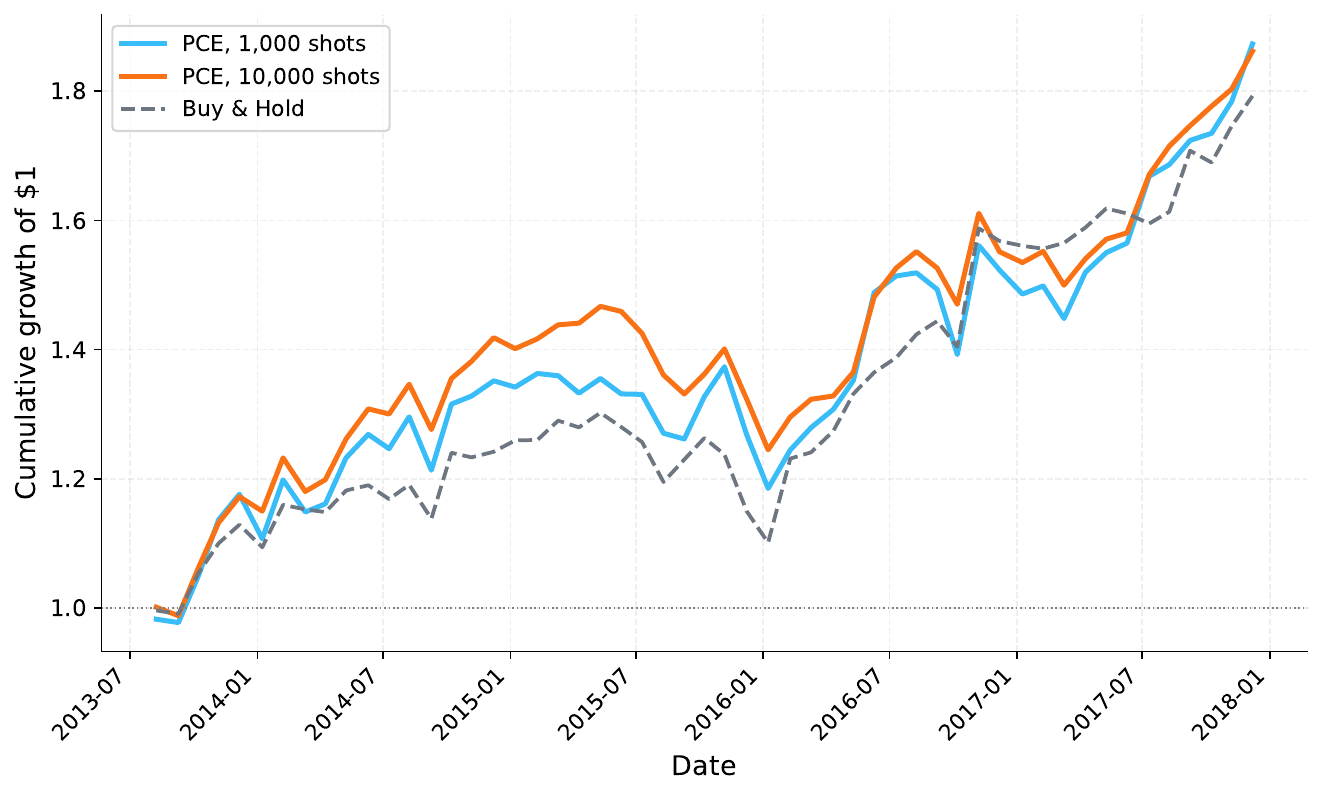}
  \caption{Streaming \pce{} at $N=1{,}601$: cumulative return, $1{,}000$ vs.\ $10{,}000$ shots.}
  \label{fig:shots-return}
\end{figure}

Whether this translates into realized portfolio return can be checked directly: at $N=1{,}601$,
running the full streaming backtest at $1{,}000$ vs.\ $10{,}000$ shots (Figure~\ref{fig:shots-return}) leaves
total return essentially unchanged ($87.3\%$ vs.\ $86.2\%$) while raising Sharpe from $4.62$ to
$5.47$ -- consistent with the cut-quality finding above: shots mostly tighten the dispersion of
outcomes around a similar mean, rather than shifting the mean return itself. A $10\times$ shot
budget increase does not buy a materially different backtest, only a smoother one.

\subsection{Scalability to large asset universes}
\label{sec:results-scalability}
Extending \S\ref{sec:results-pce-streaming}--\ref{sec:results-comparison} well beyond $N=300$:
\pce{}'s runtime grows $\sim\!13\times$ from $N=50$ to $N=300$ (0.5s$\to$6.5s per rebalance),
while \dpce{}'s grows $\sim\!111\times$ over the same range (0.65s$\to$73s) -- an order of
magnitude worse, because \dpce{}'s qubit requirement scales with $\kappa n$
(\S\ref{sec:dpce-algorithm}) while \pce{}'s recursive splits operate on shrinking subgraphs. This
is why every result beyond $N=300$ below uses \pce{} only.

\begin{figure}[t]
  \centering
  \includegraphics[width=\linewidth]{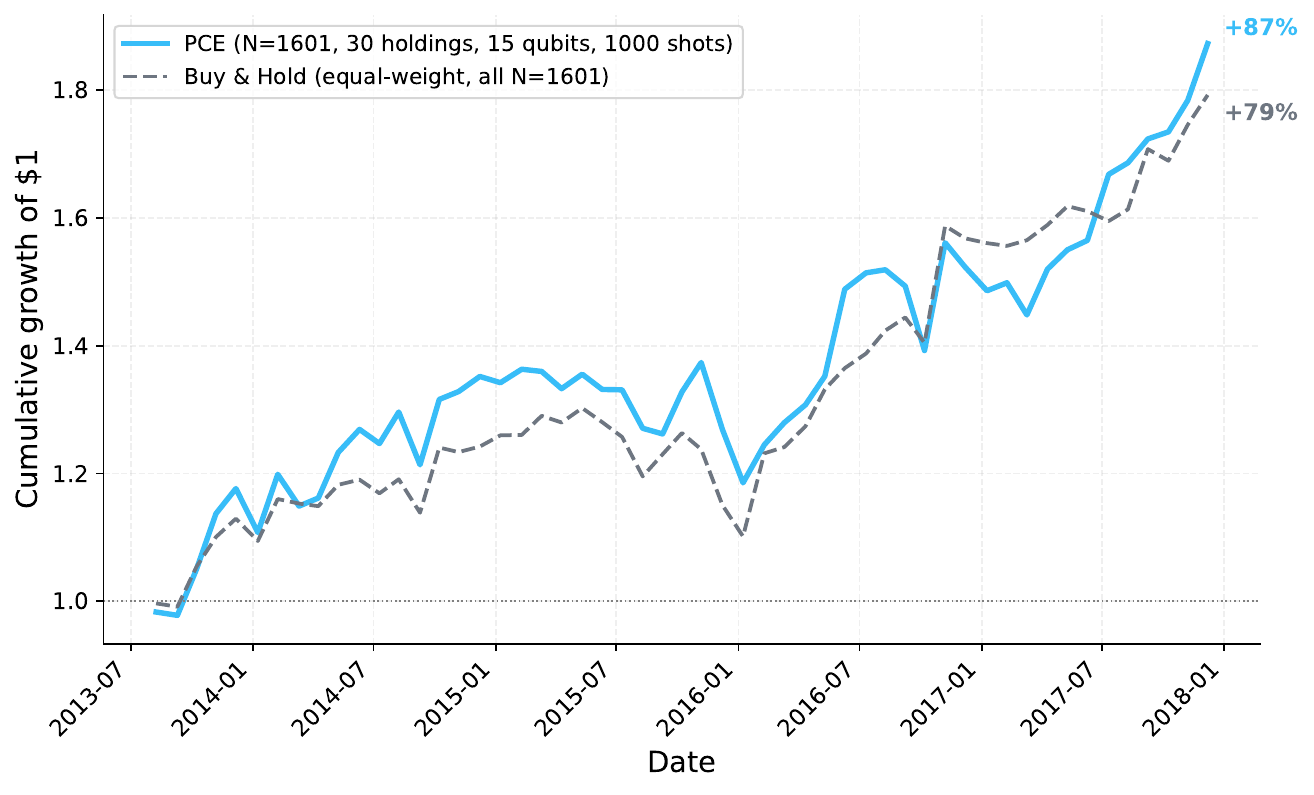}
  \caption{\pce{} vs.\ Buy \& Hold, $N=1{,}601$, same 2013--2018 window as every smaller $N$
  above.}
  \label{fig:n1601}
\end{figure}

At $N=1{,}601$ -- a $5.3\times$ increase over the largest universe size tested in
\S\ref{sec:results-pce-streaming}--\ref{sec:results-comparison} -- \pce{} still beats
Buy~\&~Hold on total return ($+87.3\%$ vs.\ $+79.2\%$, Table~\ref{tab:scalability-full},
Figure~\ref{fig:n1601}), confirming the streaming deployment continues to add value at a scale
an order of magnitude beyond what a static study could establish. The margin is narrower than at
$N=50$ ($8$ points vs.\ $254$), and Sharpe at this $N$ slightly favors Buy~\&~Hold ($5.03$ vs.\
$4.62$) -- a reminder that return and risk-adjusted return need not move together as $N$ grows,
and worth tracking as a trend rather than treating the $N=50$--$300$ result as scale-invariant.

\begin{figure}[t]
  \centering
  \includegraphics[width=\linewidth]{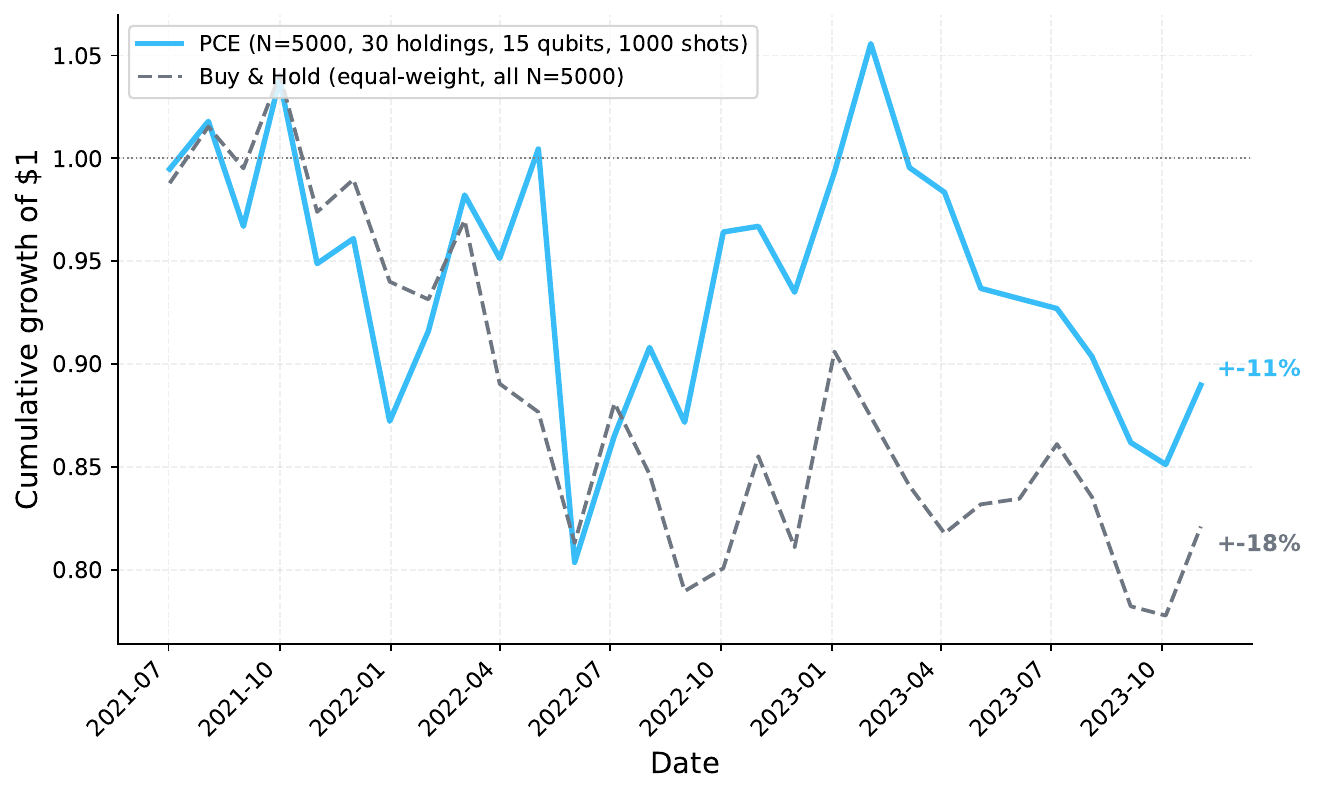}
  \caption{\pce{} vs.\ Buy \& Hold, $N=5{,}000$, 2021--2024 window (different period from every
  other result in this section).}
  \label{fig:n5000}
\end{figure}

A further point at $N=5{,}000$ (\texttt{order}$=4$, $15$ qubits, $n_{\text{clusters}}=30$,
$1{,}000$ shots) is reported separately rather than as a continuation of this trend, because
reaching 5,000 clean tickers (\S\ref{sec:data}) is only possible on a different, shorter
2021--2024 window, not the 2013--2018 window shared by every $N$ above. Over that period
-- which spans the 2022 growth and technology drawdown -- both \pce{} and Buy~\&~Hold post
negative total returns, and \pce{} loses less ($-11.0\%$ vs.\ $-17.9\%$; Sharpe $-0.47$ vs.\
$-1.76$, Table~\ref{tab:scalability-full}, Figure~\ref{fig:n5000}). This result is informative on
its own terms -- the streaming deployment continues to outperform the passive benchmark at
$N=5{,}000$, in a market regime the $N\leq1{,}601$ results do not cover at all -- but window and
scale are confounded in this single data point, so it should not be read as direct evidence about
how the $N=1{,}601$ margin would continue to narrow or widen at $N=5{,}000$ on the original
window.

The two windows used in this paper are, in effect, a pre-pandemic and a post-pandemic test: every
$N\leq1{,}601$ result above uses 2013--2018, a sustained pre-COVID bull market in which both
\pce{} and Buy~\&~Hold post large positive returns at every $N$ (Table~\ref{tab:scalability-full}),
while the $N=5{,}000$ window (2021--2024) begins just after the 2020--2021 pandemic recovery rally
and runs directly through the 2022 rate-hike-driven growth and technology drawdown -- a period in
which equity markets broadly declined rather than merely being more volatile. This is the correct
regime to ask whether \pce{} only adds value when the underlying market is rising, or whether the
diversification it constructs also limits losses when it is not. The rebalance-level detail
behind Figure~\ref{fig:n5000} answers this directly: \pce{}'s worst point over the full backtest
is a trough of $0.804$ per dollar invested ($-19.6\%$) on 2022-06-02, against Buy~\&~Hold's worse
trough of $0.778$ ($-22.2\%$) on 2023-10-04 -- \pce{} loses less at its single worst point, not
only on average. More strikingly, \pce{} recovers to a positive cumulative return (above the
original \$1) at $4$ of the $29$ rebalance dates in this window, peaking at $+5.6\%$ on
2023-02-02, while Buy~\&~Hold is above water at only $2$ of $29$ dates and never again after
2021-10-01, less than three months into the backtest. So while the headline $N=5{,}000$ total
return is negative for both strategies, \pce{}'s diversified, re-clustered portfolio is not
simply ``less negative'' in aggregate -- at several points well into the drawdown it is genuinely
ahead of its own starting capital, something the passive benchmark never again achieves over the
same period.

\begin{table}[t]
\centering
\caption{Scalability progression. $N=50$--$1{,}601$ share the identical 2013--2018 window;
$N=5{,}000$ uses a different, more recent window (see text) and is not directly comparable.}
\label{tab:scalability-full}
\small
\resizebox{\linewidth}{!}{%
\begin{tabular}{lrrrr}
\toprule
$N$ & PCE return & B\&H return & PCE Sharpe & BH Sharpe \\
\midrule
50      & 338.2\% & 83.9\% & 7.94 & 5.09 \\
300     & 125.1\% & 66.4\% & 6.68 & 5.37 \\
1{,}601 & 87.3\%  & 79.2\% & 4.62 & 5.03 \\
\midrule
5{,}000$^\dagger$ & $-$11.0\% & $-$17.9\% & $-$0.47 & $-$1.76 \\
\bottomrule
\end{tabular}}
\\[2pt]
\raggedright\footnotesize $^\dagger$Different (2021--2024) backtest window; not directly
comparable to the rows above.
\end{table}

%% file: sections/05_conclusion.tex
This paper asked whether \pce{}'s static, whole-history promise~\citep{soloviev2026pce} survives
the move to a live, rolling deployment, and the answer is yes, up to a scale an order of
magnitude beyond what that static study or a one-qubit-per-asset encoding could reach. Across
every tested combination of universe size ($N=50$--$300$), correlation window, risk-aversion
level, and rebalance cadence, both streaming variants introduced here -- \pce{}'s iterative
recursive bipartition and \dpce{}'s joint multi-cut -- beat an equal-weight Buy~\&~Hold baseline
without exception. That advantage persists at $N=1{,}601$, a $5.3\times$ increase over the $N=300$
ceiling of the smaller-scale experiments, confirming that \pce{}'s $O(\log n)$ qubit compression
is not a static-snapshot artifact: it continues to add value once the correlation graph,
clustering, and portfolio selection are all re-run on a rolling clock, with the compounding cost
and benefit that implies.

The single most important result in this paper, however, is not the bull-market numbers above --
it is what happens at $N=5{,}000$ on the 2021--2024 window, which spans the post-pandemic
growth-and-technology drawdown rather than the sustained bull market every smaller-$N$ result
uses. A quantum-inspired portfolio method that only adds value when markets are rising would be a
curiosity, not a deployment case. It does not: \pce{} loses less than Buy~\&~Hold on total return
($-11.0\%$ vs.\ $-17.9\%$) and at its single worst point ($-19.6\%$ vs.\ $-22.2\%$), and crosses
back above its own starting capital at several rebalances where Buy~\&~Hold never recovers after
the first few months of the window. Losing less than a declining baseline is itself the result,
not a consolation beside the bull-market figures -- it is direct evidence that the diversification
the graph-cutting objective constructs earns its keep precisely in the regime where capital
preservation matters most, which the pre-pandemic window cannot demonstrate at all. We report
this as one data point, with scale and market regime confounded (\S\ref{sec:results-scalability}),
not as a second clean trend alongside $N=50$--$1{,}601$ -- but it is the result that most directly
answers whether this approach is useful beyond a rising market, and it says yes.

\pce{} and \dpce{} are not interchangeable in practice, despite tracking each other closely in
raw return. \dpce{}'s single joint call needs strictly more qubits than \pce{}'s largest split
whenever more than two clusters are requested, and its runtime scales roughly an order of
magnitude worse with $N$ (\S\ref{sec:results-scalability}); it is also close to twice as sensitive
to rebalancing cadence as \pce{} is, and that sensitivity grows with $N$ and
$n_{\text{clusters}}$ (\S\ref{sec:results-rebalance-vs-risk-aversion}). \dpce{} is the better
choice only when both conditions needed to realize its performance -- a small-to-moderate
universe and a short rebalancing step -- are simultaneously affordable; \pce{} is the safer
default whenever either constraint binds, which is why it is the only algorithm carried to
$N>300$ in this paper.

Among the operational parameters a streaming deployment must actually choose rather than fix
once, rebalance cadence is the cleaner control: \texttt{step\_size} moves turnover and total
return monotonically in the same direction (\S\ref{sec:results-rebalance-vs-risk-aversion}),
while the risk-aversion dial $\rho$ moves three coupled mechanisms at once and lands
non-monotonically on Sharpe (\S\ref{sec:results-risk-aversion}) -- useful as a profile selector
among already-favorable outcomes, not as a risk-reduction control in the textbook sense. The
$126$-day correlation window is the consistent best performer across every universe size tested,
but we could not fully decompose why: a Marchenko--Pastur signal-fraction diagnostic
(\S\ref{sec:results-pce-streaming}) confirms longer windows give a statistically cleaner
correlation estimate, yet that alone predicts the longest window should perform best, not worst --
the likely explanation is a staleness-versus-noise trade-off that the recency-weighting mechanism
of \S\ref{sec:method-window} only partially addresses, and we leave a clean decomposition to
future work. Warm-starting raises solution quality without speeding up convergence
(\S\ref{sec:results-warmstart}), $1{,}000$ measurement shots already saturate realized portfolio
return at $N=1{,}601$ with further shots only tightening dispersion around the same mean
(\S\ref{sec:results-shots}), and transaction-cost-aware selection cuts turnover by $11.3\%$ on
average with no detectable cost to the (frictionless) backtest return
(\S\ref{sec:results-tx}) -- a turnover reduction that, charged against any realistic real-world
commission rate, is a genuine efficiency gain rather than a trade-off against return.

Four limitations follow from how this study is scoped; the first two bear most directly on how
the headline numbers above should be read. First, every comparison is against equal-weight
Buy~\&~Hold over the \emph{full} candidate universe, not against a baseline matched in holding
count: a portfolio of $k$ names selected uniformly at random from the same universe would likely
also outperform a full-universe equal-weight benchmark in many regimes, for reasons unrelated to
cluster quality (concentration and idiosyncratic-volatility effects alone). This study has no
random-subset-of-size-$k$ control, so the results above are evidence that \pce{}/\dpce{}-selected
portfolios outperform the realistic passive alternative, not isolated evidence that graph-cutting
selection specifically outperforms an equally-concentrated but arbitrary one -- that decomposition
is the most informative missing experiment and a natural next step. Second, every number in this
paper is a point estimate from a single historical path per configuration: we report no bootstrap
confidence intervals or significance tests on any margin, including the $N=1{,}601$ return gap and
the rank-correlation figures computed from as few as four configurations
(\S\ref{sec:results-rebalance-vs-risk-aversion}); the direction of every effect reported is
consistent across the configurations tested, but its statistical reliability under resampling is
not established here. Third, the $N=5{,}000$ result is one data point in which scale and market
regime change simultaneously; isolating the two would need either a controlled $N=5{,}000$ run on
a 2013--2018-style window or a longer, cleaner post-pandemic dataset at the smaller universe
sizes, neither of which was available here. Fourth, \dpce{}'s growing sensitivity to rebalancing
cadence as $N$ grows was only characterized up to $N=100$; whether it continues to worsen,
plateaus, or reverses at the universe sizes where \pce{} alone was evaluated ($N$ up to
$5{,}000$) is open. All four are tractable extensions of the pipeline already built here, not
changes to it. What this
paper establishes is narrower and, we think, more load-bearing: \pce{}'s qubit-efficient encoding
is not only a static-benchmark result but a property that survives the realistic friction of a
live, repeatedly-rebalanced deployment, including in the one market regime -- a sustained
downturn -- where a diversification-based method most needs to prove it is not merely riding a
rising tide.

%% file: sections/06_appendix_datasets.tex
\section{Datasets}
\label{app:data}

This appendix describes the three daily-OHLCV equity datasets used in \S\ref{sec:results} and
reports descriptive statistics for each, computed directly from the filtered price series
actually consumed by the corresponding experiment (\S\ref{sec:data}), not from the raw source
file.

\subsection{Data sources and filtering criteria}
Table~\ref{tab:appendix-datasets} summarizes the three sources. In every case the backtest
window is fixed first, then only tickers with a complete (gap-free) daily close over that window
are retained, and any ticker with a single-day move exceeding $50\%$ -- even in
split/dividend-adjusted prices -- is dropped as a data-quality check (\S\ref{sec:data}); the
table reports ticker counts before and after this filtering. The $N=50$--$300$ results and the
market-graph example of \S\ref{sec:market-graph-example} use the S\&P~500 dataset; the
$N=1{,}601$ result (\S\ref{sec:results-scalability}) uses the \texttt{prices.csv} dataset
restricted to the identical 2013--2018 window; the $N=5{,}000$ result uses a 2021--2024 window
of the \texttt{all\_stock\_data.csv} dataset, the only one of the three with enough simultaneous
gap-free tickers to reach $N=5{,}000$.

\begin{table}[h]
\centering
\caption{The three datasets underlying \S\ref{sec:results}.}
\label{tab:appendix-datasets}
\small
\resizebox{\linewidth}{!}{%
\begin{tabular}{lrrrr}
\toprule
Dataset & Raw tickers & Window & Complete & Clean ($N$) \\
\midrule
S\&P 500, 5yr        & 505   & 2013-02-08--2018-02-07 & 470   & 470 (300 used) \\
\texttt{prices.csv}        & 2{,}070 & 2013-02-08--2018-02-07 & 1{,}671 & 1{,}601 \\
\texttt{all\_stock\_data.csv} & 9{,}315 & 2021-01-01--2024-01-01 & 6{,}869 & 5{,}049 (5{,}000 used) \\
\bottomrule
\end{tabular}}
\end{table}

\subsection{Pairwise return correlation}
\begin{figure}[h]
  \centering
  \includegraphics[width=\linewidth]{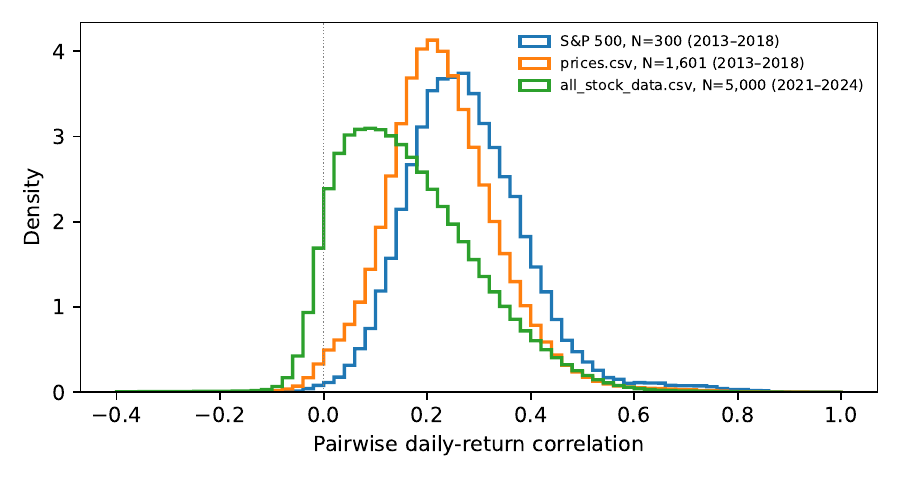}
  \caption{Distribution of pairwise daily-return correlation $\hat\rho_{uv}$ within the clean
  universe used for each of the three results, computed over the exact window and ticker set
  consumed by that experiment.}
  \label{fig:appendix-corr}
\end{figure}

Table~\ref{tab:appendix-corr} reports summary statistics of the pairwise daily-return
correlation $\hat\rho_{uv}$ (Eq.~\eqref{eq:edge-weight}'s input) across all ticker pairs in each
clean universe, and Figure~\ref{fig:appendix-corr} shows the full distributions. All three are
unimodal and centered well above zero, consistent with the common-factor structure expected of
equity returns, but the center and spread differ with universe composition and period: the
$N=300$ S\&P~500 subset -- all large-cap, all in the same index, same five-year window -- has
the highest mean pairwise correlation ($0.278$) and the thinnest negative tail ($0.3\%$ of pairs
below zero); the broader $N=1{,}601$ universe over the identical window is less uniformly
correlated (mean $0.230$, $1.3\%$ negative), consistent with it spanning a wider mix of sectors
and market capitalizations than the S\&P~500 alone; and the $N=5{,}000$ universe, over the
different and more turbulent 2021--2024 window, has both the lowest mean correlation ($0.166$)
and the heaviest negative tail ($7.0\%$ of pairs), which is itself a useful cross-check on
\S\ref{sec:results-scalability}'s reported negative return at that point -- it is a sample drawn
from a genuinely more adverse and more cross-sectionally dispersed regime, not only a larger one.
In all three cases the graph-cutting construction of \S\ref{sec:finance-mapping} still has
contrast to exploit: even the lowest-mean-correlation universe has $1.5\%$ of pairs above
$\hat\rho=0.5$ and a non-trivial negative-correlation tail, both of which the edge-weight
transformation of Eq.~\eqref{eq:edge-weight} is designed to separate into different clusters.

\begin{table}[h]
\centering
\caption{Pairwise correlation statistics, clean universe, per dataset.}
\label{tab:appendix-corr}
\small
\resizebox{\linewidth}{!}{%
\begin{tabular}{lrrrrrr}
\toprule
Universe & Mean & Median & Std & \% $<0$ & \% $>0.5$ & Pairs \\
\midrule
S\&P 500, $N{=}300$          & 0.278 & 0.267 & 0.118 & 0.3 & 3.8 & 44{,}850 \\
\texttt{prices.csv}, $N{=}1{,}601$ & 0.230 & 0.222 & 0.111 & 1.3 & 1.7 & 1{,}280{,}800 \\
\texttt{all\_stock\_data.csv}, $N{=}5{,}000$ & 0.166 & 0.147 & 0.134 & 7.0 & 1.5 & 12{,}497{,}500 \\
\bottomrule
\end{tabular}}
\end{table}

